\documentclass[twocolumns]{aa}  

\usepackage{graphicx}
\usepackage{txfonts}
\begin{document}

    \title{The structural properties of multiple populations in globular clusters: the instructive case of NGC~3201}


     \author{Mario~Cadelano \inst{1,2}
          \and
          Emanuele~Dalessandro \inst{2}
          \and
          Enrico Vesperini\inst{3}
          }

   \institute{Dipartimento di Fisica e Astronomia “Augusto Righi'', Universit\`a degli Studi di Bologna, Via Gobetti 93/2, 40129 Bologna, Italy \\
         \and
             INAF-Osservatorio di Astrofisica e Scienze dello Spazio di Bologna, Via Gobetti 93/3 I-40129 Bologna, Italy \\
             \and
             Department of Astronomy, Indiana University,Swain West, 727 E. 3rd Street, IN 47405 Bloomington, USA
             }

   \date{Received September 15, 1996; accepted March 16, 1997}

 
  \abstract{
All multiple population (MP) formation models in globular clusters (GCs) predict that second population (SP) stars form more centrally concentrated than the first population (FP). As dynamical
evolution proceeds, spatial differences are progressively erased, and only dynamically young clusters are expected to still retain a partial memory of the initial structural differences. In recent years, this picture has been supported by observations of the MP radial distributions of both Galactic and extragalactic GCs. However, recent observations have suggested that in some systems, FPs might actually form more centrally segregated, with NGC~3201 being one significant example of such a possibility. Here we present a detailed
morphological and kinematic characterization of the MPs in NGC~3201 based on a combination of photometric and astrometric data. We show that the distribution of the SP is clearly bimodal.
Specifically, the SP is significantly more centrally concentrated than the FP within $\sim1.3$ cluster’s half-mass radius. Beyond this point, the SP fraction increases again, likely due to asymmetries in the spatial distributions of the two populations. The central concentration of the SP observed in the central regions implies that it formed more centrally concentrated than the FP, even more so than what is observed in the present-day. This interpretation is supported by the key information provided by the MP kinematic properties. Indeed, we find that the FP is isotropic across all the sampled cluster extension, while the velocity distribution of the SP becomes radially anisotropic in the cluster’s outer regions, as expected for the dynamical evolution of SP stars formed more centrally concentrated than the FP. The combination of spatial and kinematic observations provide key insights into the dynamical properties of this cluster and lend further support to scenarios in which the SP forms more centrally concentrated than the FP.}

    
    \titlerunning{Multiple populations in NGC~3201}
   \maketitle

%
\section{Introduction} \label{sec:intro}
The presence of multiple populations (MPs) differing in terms of light-element abundances (e.g. He, C, N, O, Na, Mg, Al) while having the same iron-peak abundance is a key property of globular clusters (GCs; see e.g., \citealt{Bastian2018}, \citealt{gratton19}). In fact, MPs are observed in nearly all relatively massive ($M>10^4 M_{\odot}$; \citealt{Carretta2010}) stellar clusters, both in the Milky Way and in external galaxies (e.g. \citealt{mucciarelli08,Dalessandro2016,larsen14,sills19,cadelano23}) and in clusters of all ages at least down to $\sim$ 2 Gyr \citep{Martocchia2018,cadelano2022}. Stars sharing the same light-element chemistry as field stars (i.e. Na-poor/O-rich, CN-weak) are classified as first-population (FP), while Na-rich/O-poor, CN-strong stars are referred to as second-population (SP). MPs are believed to form during the very early epochs of GC life ($<100$ Myr). A number of theoretical studies have been put forward over the years, but no consensus has been reached yet on their origin. The most popular scenarios invoke formation of SP stars out of a mix of pristine gas and 
processed gas ejected by AGB stars, massive binary stars, or super massive stars forming through runaway collisions during the 
cluster early lifetime (e.g.  \citealt{Decressin2007a,dercole08,demink09,Bastian2013,Denissenkov2014,dantona16,gieles18,renzini22,winter23}). 

The morphological and kinematical properties of MPs are powerful tools to constrain their formation and evolution scenarios. 
The majority of the formation models predicts that the SP forms more centrally segregated and, 
possibly, more rapidly rotating than a more spatially extended FP population 
(see e.g. \citealt{dercole08,calura19,Bekki2011,lacchin22}).
Differences among the morphological and kinematical properties of MPs are progressively erased during GC long-term dynamical evolution (see e.g. \citealt{Vesperini2013,Henault-Brunet2015,miholics15,Dalessandro2018a,dalessandro2021,Bellini2015}), but dynamically younger clusters are expected to retain some memory of their primordial differences. 

In \citet{Dalessandro2019}, we studied the radial distributions of MPs in a sample of 20 GCs spanning a broad range of dynamical ages. The relative degree of segregation between FP and SP stars was quantitatively measured by means of the $A^+$ parameter, defined as the area enclosed between their cumulative radial distributions within twice the cluster half-light radii ($r_{hl}$). Our study has revealed a clear trend between $A^+$ and GC degree of internal dynamical evolution, as constrained by the ratio of the cluster’s age to its half-mass relaxation timescale ($t/t_{rh}$). Less dynamically evolved clusters ($t/t_{rh} < 8 - 10$) have SP stars more centrally concentrated than FPs (i.e. negative values of $A^+$), while in  dynamically evolved systems the relative differences between FP and SP stars decrease and eventually disappear ($A^+$ tends to zero). Such a behavior is broadly consistent with predictions by $N$-body and Monte Carlo simulations following the long-term dynamical evolution of MPs \citep{Vesperini2013,Vesperini2018,Dalessandro2018c,Vesperini2021,Sollima2021} in clusters forming with an initially more centrally concentrated SP sub-system.
In a recent study,  \citet{leitinger2023} analyzed a sample of 28 GCs using a combination of HST and ground-based observations and measuring the $A^+$ parameter within $\sim4 r_{hl}$. 
In that study, the authors found that dynamically old clusters show, as expected, mixed populations, 
while clusters with $t/t_{rh}<4$\footnote{Notice that in \citet{Dalessandro2019} the values of $t_{rh}$ are those from the Harris catalogue while in \citet{leitinger2023} the value from the Baumgardt's catalogue are adopted. Besides differences in the values of the masses and structural parameters and the numerical constant in the Coulomb logarithm, the main difference between the two estimates are that the values of $t_{rh}$ in the Harris catalogue are calculated using the observed 2D half-light radius while in the Baumgardt's catalogue the 3D half-mass radius is used.} can attain any $A^+$ value (i.e., both positive and negative values 
along with values close to zero) and they suggested this is evidence that MPs in GCs can form with any initial relative concentration. 
While, as noted above, dynamically young clusters are indeed expected to retain some memory of the initial differences, 
it is important to emphasize that clusters undergo a rapid phase of mixing during the very early phases of their formation and evolution, followed by a more gradual mixing during their long-term evolution (see \citealt{Dalessandro2019,Vesperini2021,Sollima2021,onorato23}). This implies that in no case the present-day structural differences between FP and SP stars observed in Galactic GCs may 
reflect the actual extent of the differences emerging at the end of the formation process. 
Even in clusters with small $t/t_{rh}$ values 
both early mixing and two-body relaxation have already significantly diluted 
(and in some cases possibly erased) the initial differences between the FP and the SP sub-populations. 
Therefore observations of small values of $A^+$ in dynamically young GCs cannot be directly interpreted as the evidence that FP and SP might form already mixed. 

On the other hand, the case of GCs with positive values of $A^+$ suggesting that FP stars are more concentrated than SP stars 
definitely deserves further investigation. First, we note that significantly positive values of $A^+$ are observed in only two dynamically young clusters, namely NGC 3201 ($t/t_{rh}=3.5$) and NGC 6101 ($t/t_{rh}=1.15$).
The radial distributions shown by \citet{leitinger2023} for these two clusters, reveals that they are both characterized by a complex radial variation of the SP/(FP+SP) number ratio; specifically, their distributions are clearly bimodal and show a relative maximum in the cluster center (i.e. a more centrally concentrated SP), a minimum at intermediate distances and a strong increase in the cluster outer regions. This behavior is particularly evident in the case of NGC~3201. 

In this paper we present a detailed morphological and kinematical analysis of the MPs in NGC~3201 aimed at 
further exploring their differences and their possible origin. NGC~3201 is better suited for this study than NGC~6101 as it is more massive thus guaranteeing larger sample of stars, it is more metal-rich ([Fe/H]=-1.6) thus making the photometric selection of MPs more solid, and it is significantly closer to the Sun (d$=4.9$ kpc) thus enabling a more robust kinematic analysis.
The outline of the paper is the following: in Section~\ref{sec:selection} we present the adopted photometric and astrometric data-set and data analysis procedures; in Section~3 we show how the MP tagging in the different data-set was performed; the morphological and kinematical analysis of MPs is presented in Section~4 and Section~5, respectively. Finally, a summary and discussion of the main results is reported in Section~6.

\section{Datasets and analysis} 
\label{sec:selection}

\begin{figure*}[h!] 
\centering
\includegraphics[scale=0.24]{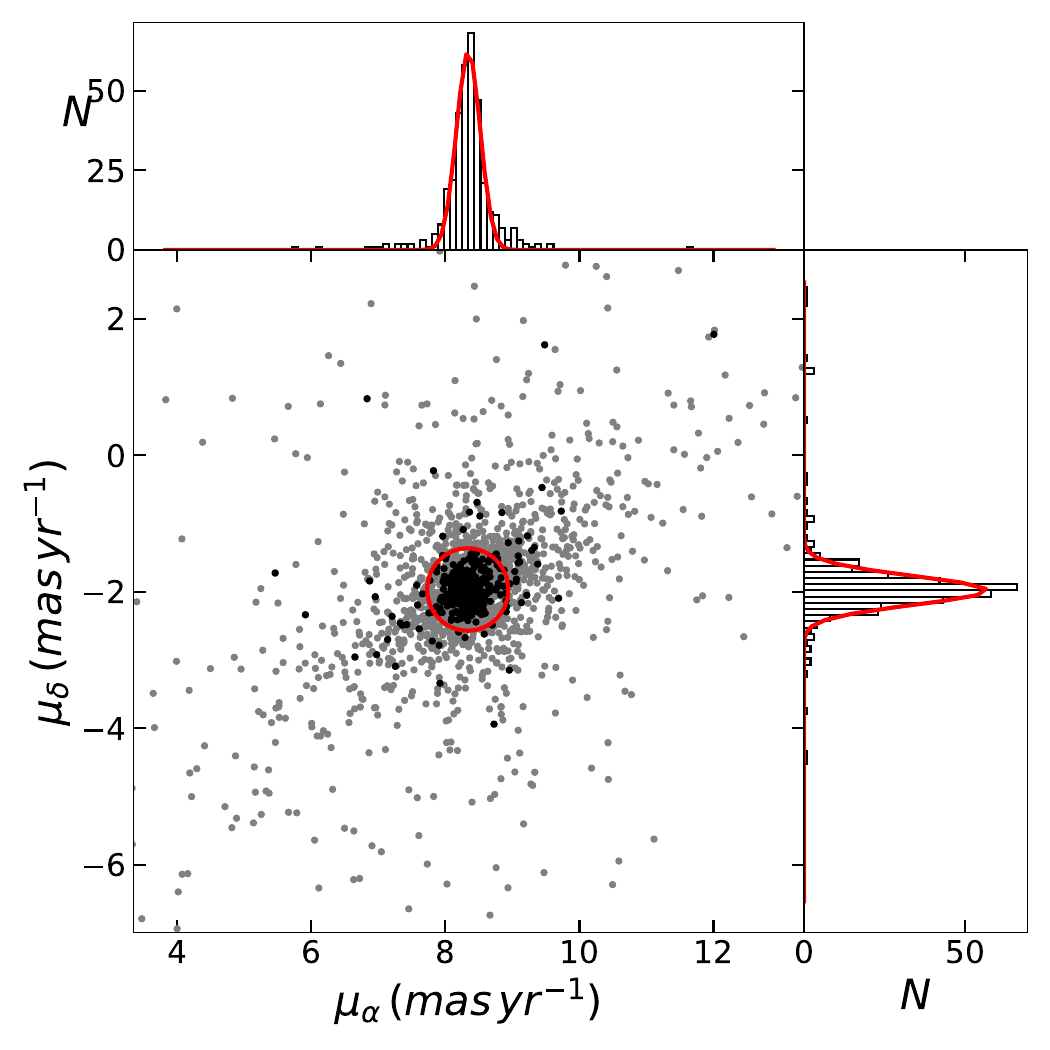}
\includegraphics[scale=0.24]{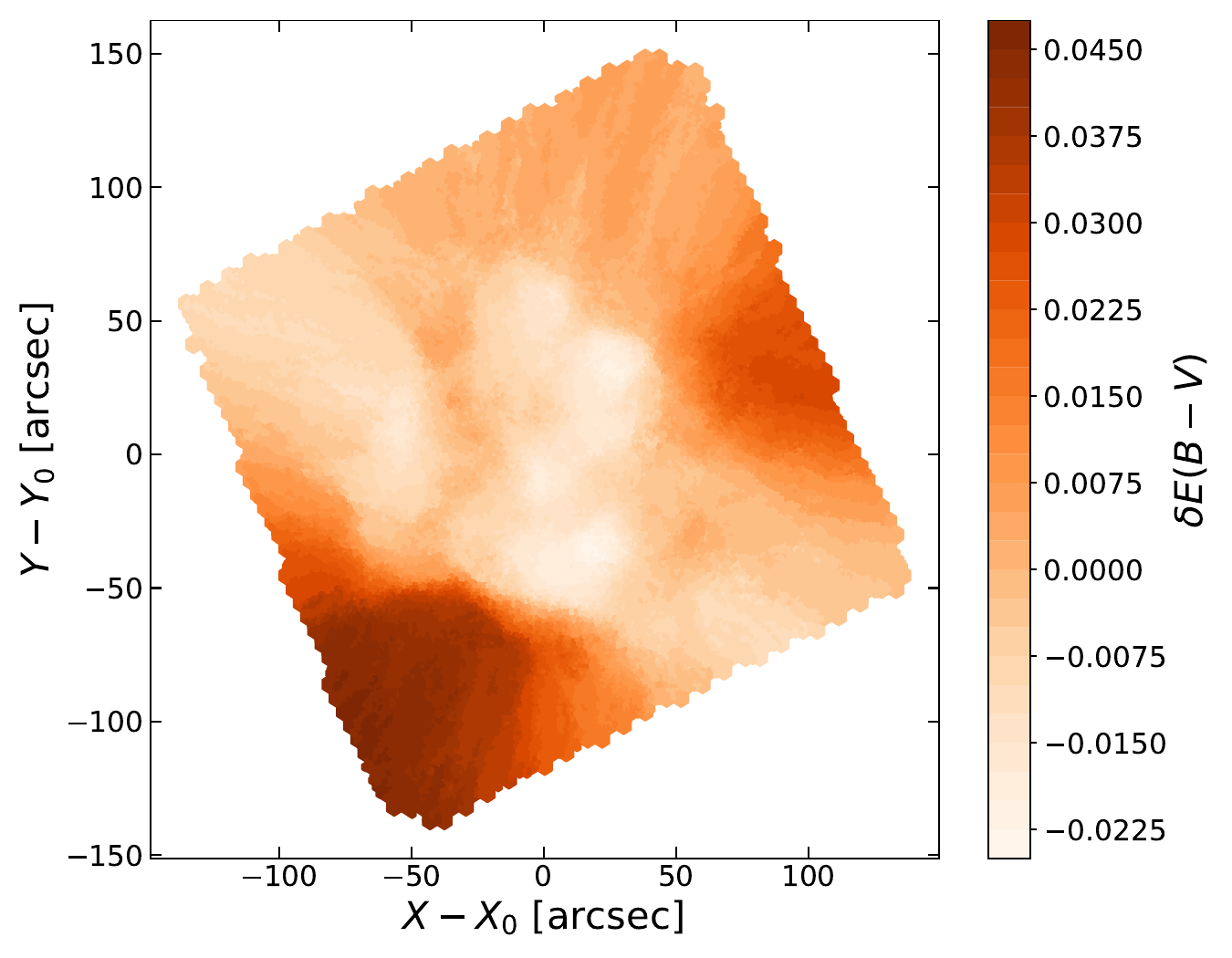}
\includegraphics[scale=0.28]{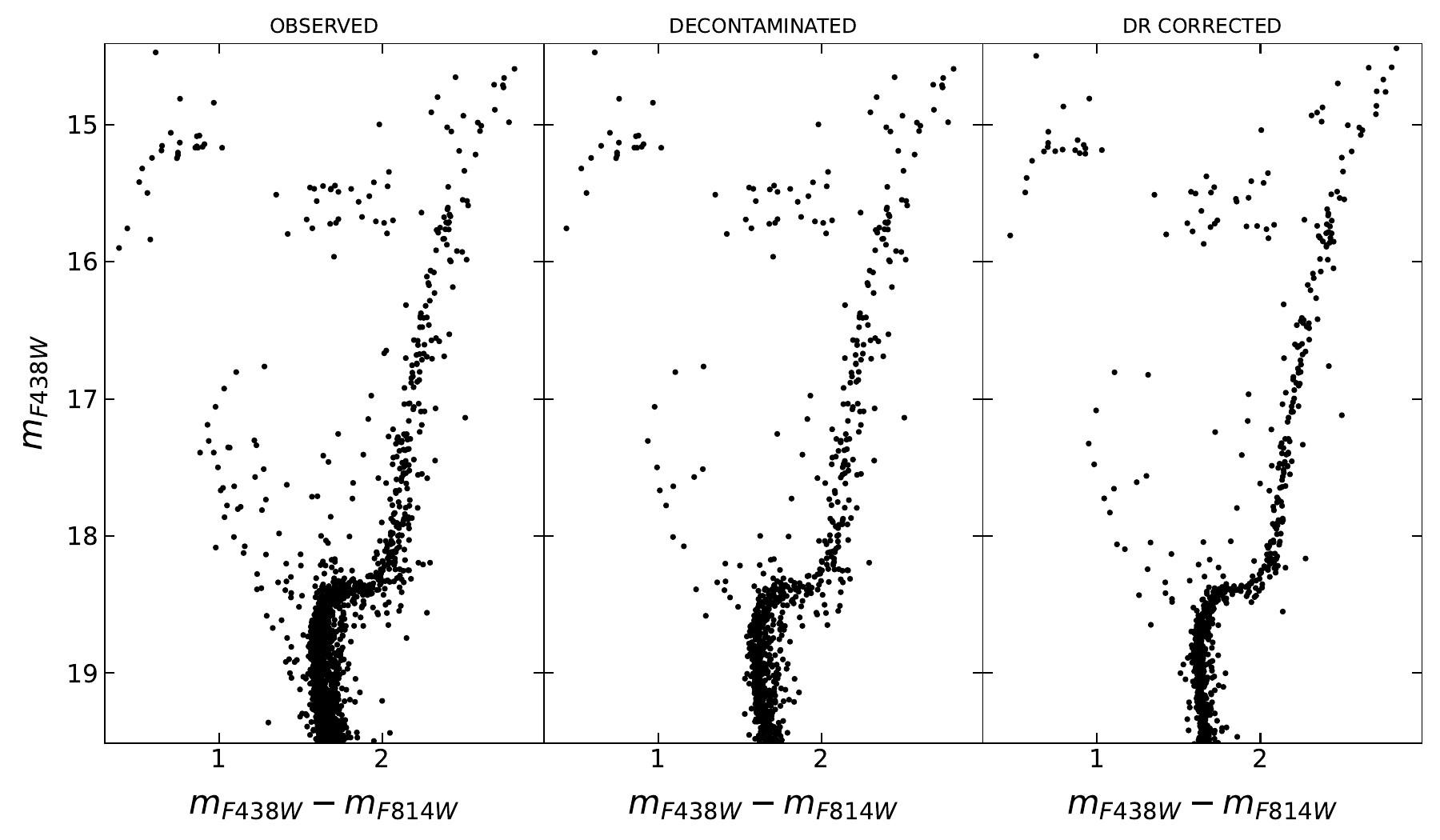}
\caption{Field decontamination and differential reddening correction of the HST data-set. {\it Top left-hand panel:} vector point diagram of the cluster stars (gray dots) as measured by Gaia DR3. RGB stars are highlighted as black dots. The upper and side panels show the histogram of the proper motions along the two axis and the best-fit Gaussian function (red curve). The red circle has a radius equal to $2\sigma$ the combined standard deviations of the two best-fit Gaussian functions. All the stars within the circle are selected as bona-fide cluster stars. {\it Top right-hand panel:} differential reddening map of NGC~3201 within the HST field of view. The color-bar on the right codifies the amount of the relative differential reddening. The coordinates along the x- and y-axes are reported with respect to the cluster center.
{\it Bottom panels:} The left-hand panel shows the ($m_{F438W}-m_{F814W}$,$m_{F438W}$) CMD of NGC~3201 including all the stars of the HST data-set. The middle and right-hand panels show the CMD following the decontamination from field interlopers and the differential reddening correction, respectively. } 
\label{fig:moti_dr_hst}
\end{figure*}

\begin{figure*}[h!] 
\centering
\includegraphics[scale=0.24]{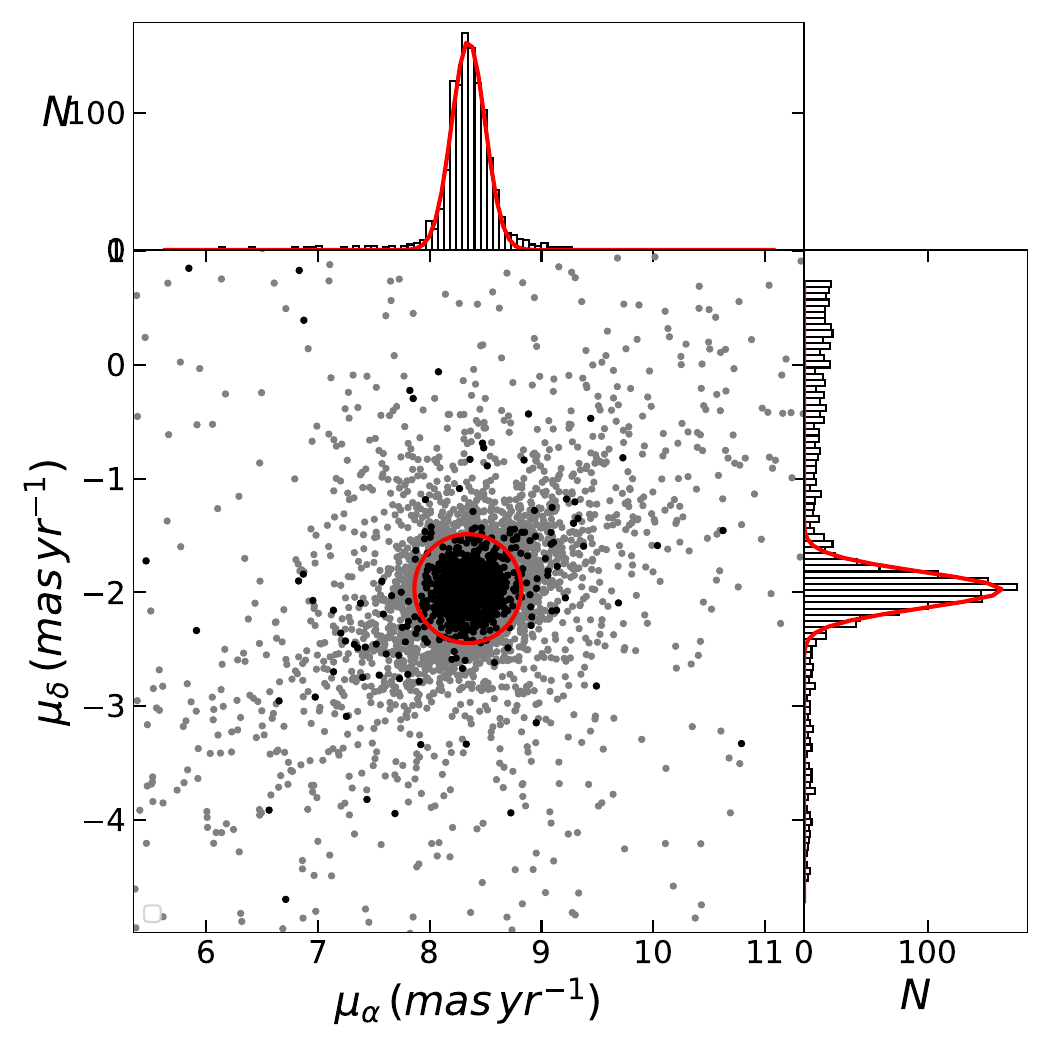}
\includegraphics[scale=0.24]{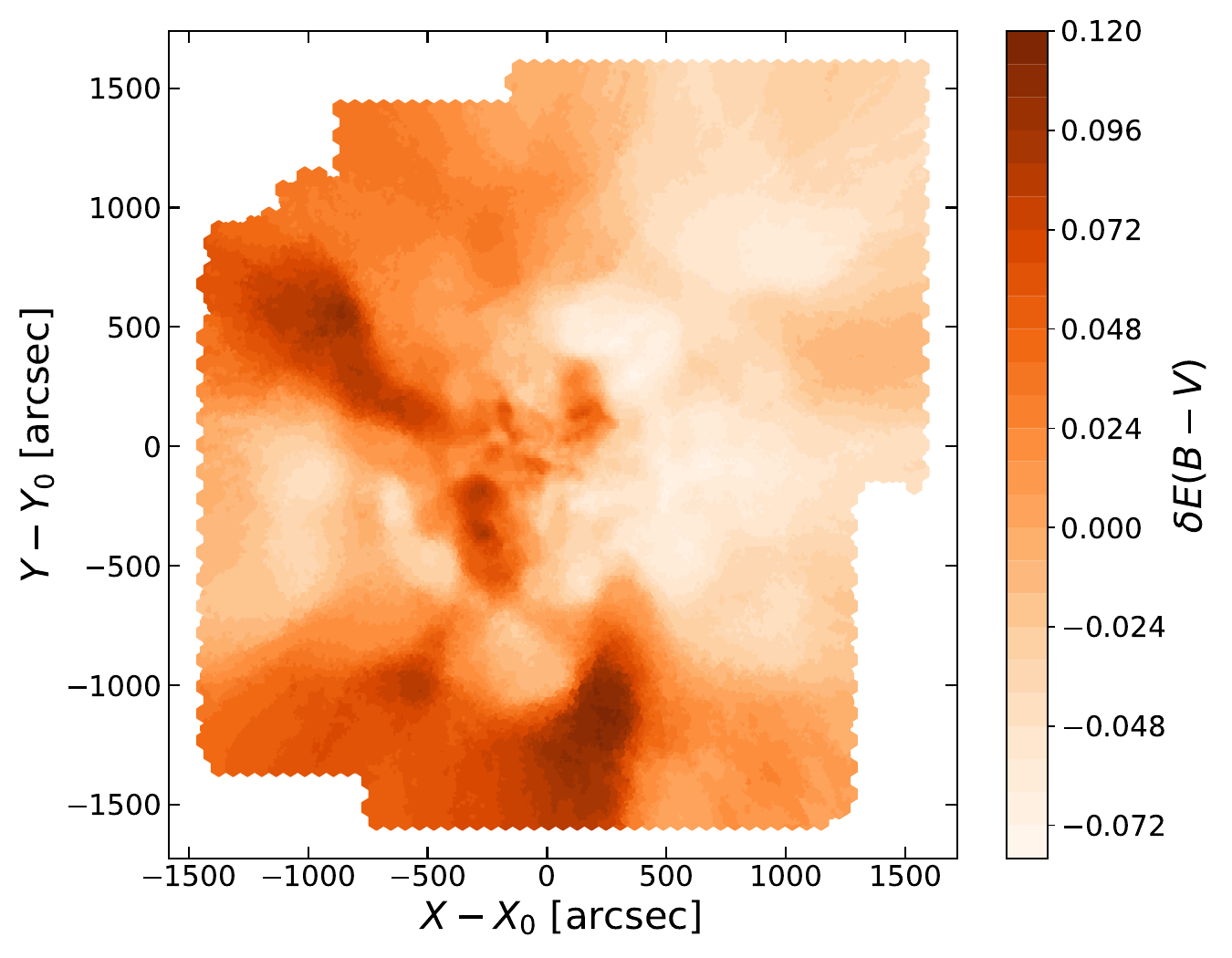}
\includegraphics[scale=0.28]{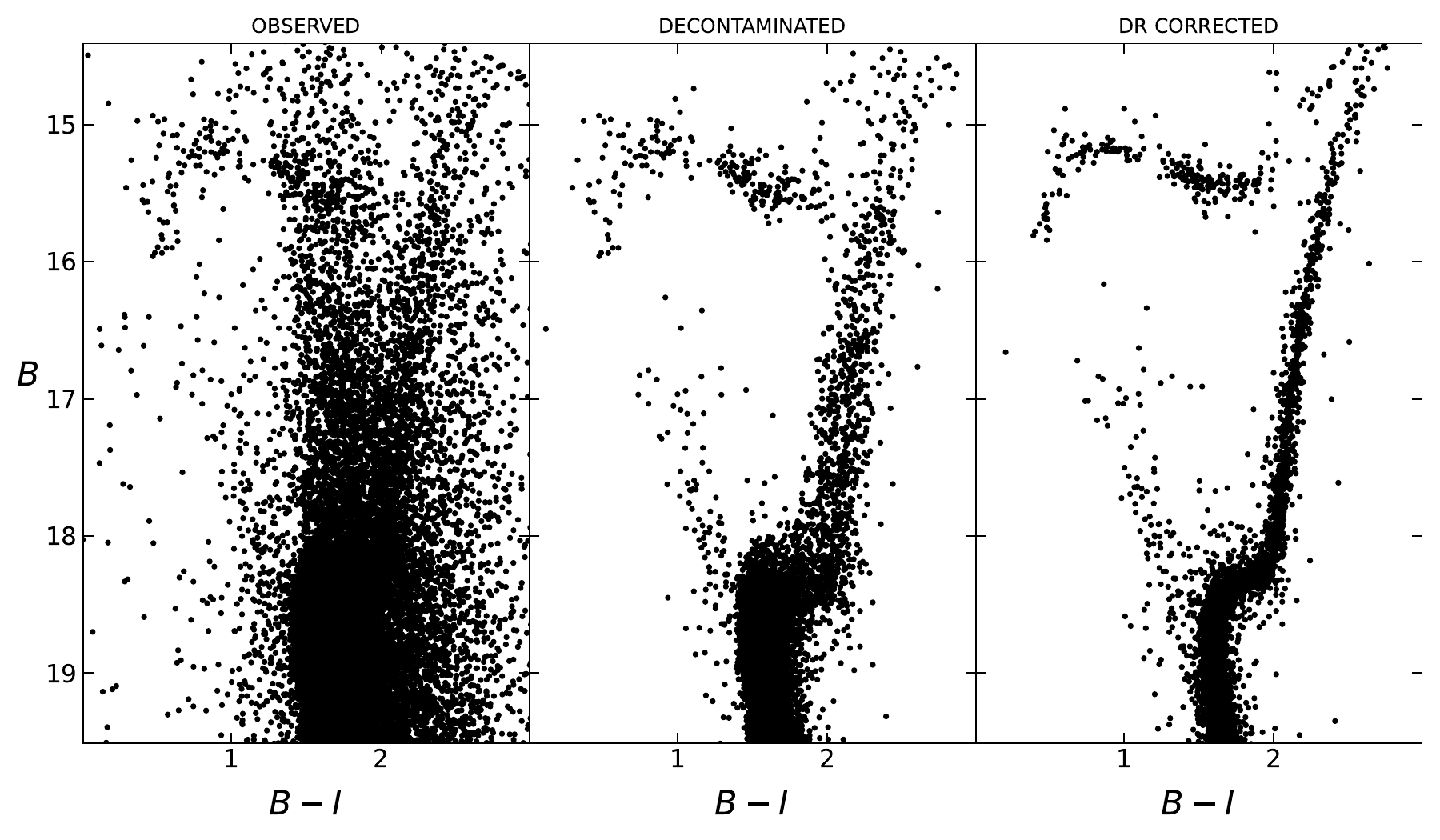}
\caption{Same as in Figure~\ref{fig:moti_dr_hst} but for the \citet{stetson2019} wide-field catalogue.} 
\label{fig:moti_dr_stetson}
\end{figure*}

This work is based on a combination of photometric and astrometric catalogues, used to tag the MPs and to perform their kinematical analysis.

We used the publicly available multi-band catalog from the HST UV Globular Cluster Survey (HUGS) \citep{piotto2015,nardiello2018} 
to sample the cluster inner region. This catalog was complemented with the UBVRI photometry from \citet{stetson2019} obtained through wide-field and ground-based observations to sample the cluster regions beyond the HST field of view. The two catalogs were cross-correlated with the Gaia DR3 \citep{gaia_dr3} star catalogue which provides the absolute proper motions of stars along with several photometric and astromeric quality indicators that will be used in the kinematical analysis. 

The adopted position of the cluster center is the one quoted by \citet{goldsbury10}: $(\alpha_0,\delta_0)=(10^h 17^m 36.82^s$,$-46^\circ 24\arcmin 44.9\arcsec$). 


As a first step, we used the Gaia proper motions to disentangle the cluster population from that of field interlopers in both the HST and wide-field data-set. To do this, we used the same method described in \citet{cadelano2020_n6256}. Briefly, we tagged as cluster members those objects having a proper motion along both the $\alpha$ and $\delta$ components within $n\sigma$ from the cluster systematic motion, where $\sigma$ comes from the best-fit Gaussian of the proper motion distributions of red giant branch stars and $n=2$. The proper motion distributions are plotted in the top left-hand panels of Figure~\ref{fig:moti_dr_hst}~and~\ref{fig:moti_dr_stetson} for the HST and wide-field catalogue, respectively. The CMD of all the stars included in the each catalogue and the corresponding ones decontaminated via proper motions are shown in the bottom left-hand and middle panel of the two figures. The comparison between the two clearly demonstrates the excellent results of the decontamination process. The following analysis will be exclusively based on the sub-sample of stars tagged as cluster members. {We stress that different proper motion selections, such as, for example, the adoption of different values of $n$, do not have a significant impact on both the morphological and kinematical results presented in the following sections.} 

The field of view sampled by the adopted catalogs is affected by significant reddening variations, which can be easily appreciated as a spread of all the evolutionary sequence along the reddening vector (see decontaminated CMDs in the bottom middle panels of Figures~\ref{fig:moti_dr_hst}~and~\ref{fig:moti_dr_stetson}). This effect can heavily hamper a proper MP tagging and thus was corrected using a technique extensively used in the past by our group \citep[e.g.][]{dalessandro2018b,cadelano20b,deras2023}. Briefly, we selected a sample of cluster stars in the I-band and F814W magnitude range from 11 to 18 and created a mean ridge line of the cluster evolutionary sequence in the $m_{F438W}-m_{F814W}$ and $B-I$ color combination for the HST and wide-field catalogs, respectively. Then we computed the distance of each one of these selected stars from the mean ridge line along the reddening vector, defined using the extinction coefficients obtained from \citet{cardelli89,girardi02}. This reference sample is used to assign a distance from the mean ridge line to all the sources in our photometric catalogs, calculated as the $\sigma$ clipped median of the distance values measured for the $n$ closest reference stars. Finally, the resulting values of the distances were easily converted into variation of the color excess $\delta E(B-V)$ using an adapted version of Equation (1) in \citet{cadelano2020_n6256}. This technique was iteratively repeated several times progressively decreasing $n$ from 80 down to 30, in steps of 10. The resulting reddening maps in the top right-hand panel of Figures~\ref{fig:moti_dr_hst}~and~\ref{fig:moti_dr_stetson} show that the sampled field of view is affected by color-excess variation up to $\delta E(B-V)\sim0.2$ mag.  The differential reddening corrected CMDs are presented in the bottom right-hand panel of the two figures and show that the adopted technique effectively removes the differential reddening effect across all the evolutionary sequences. The following analysis will be exclusively based on differential reddening corrected magnitudes.

\section{Multiple Population Tagging}
\subsection{HST Chromosome Map Selection}

\begin{figure}[] 
\centering
\includegraphics[scale=0.28]{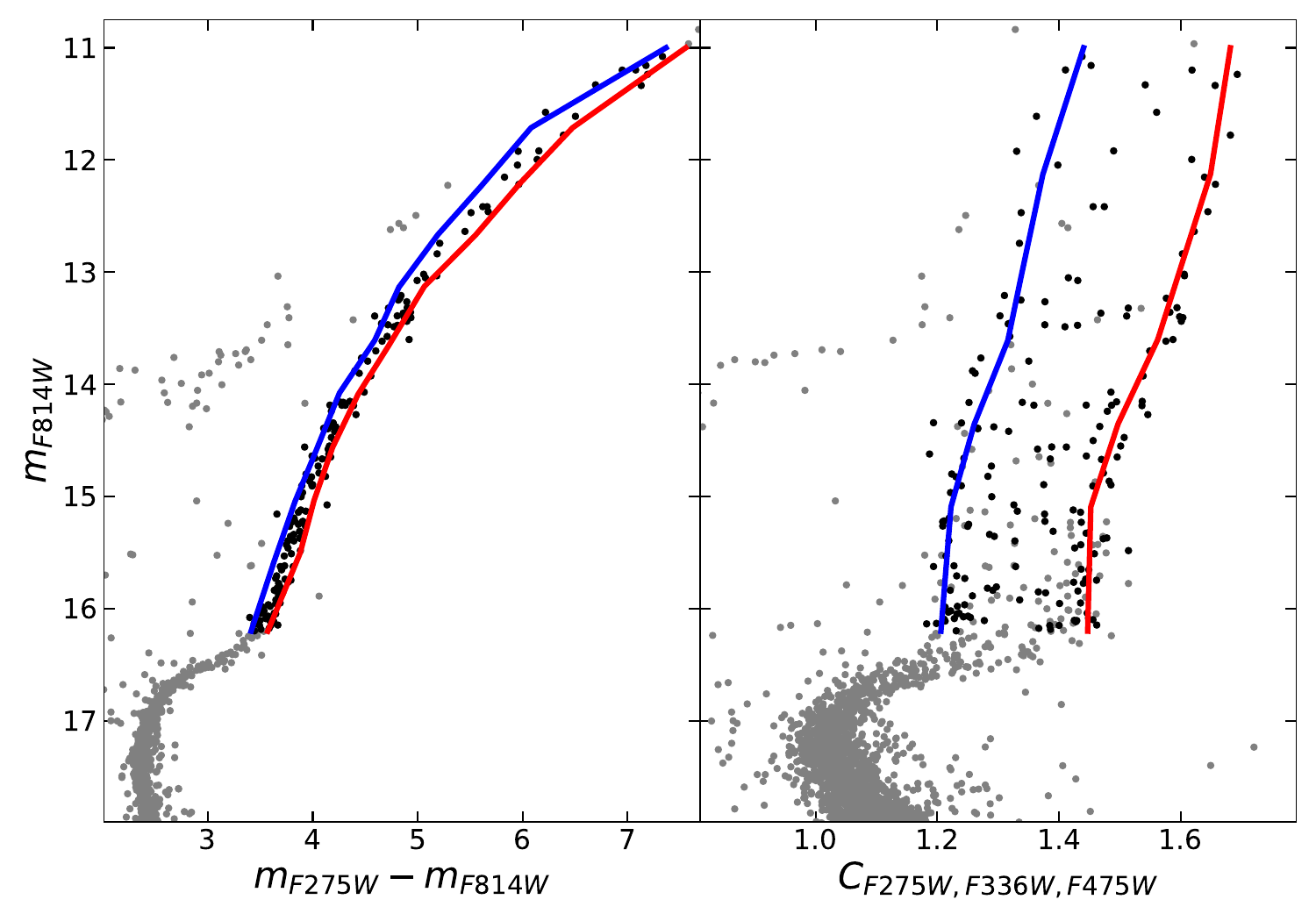}
\includegraphics[scale=0.28]{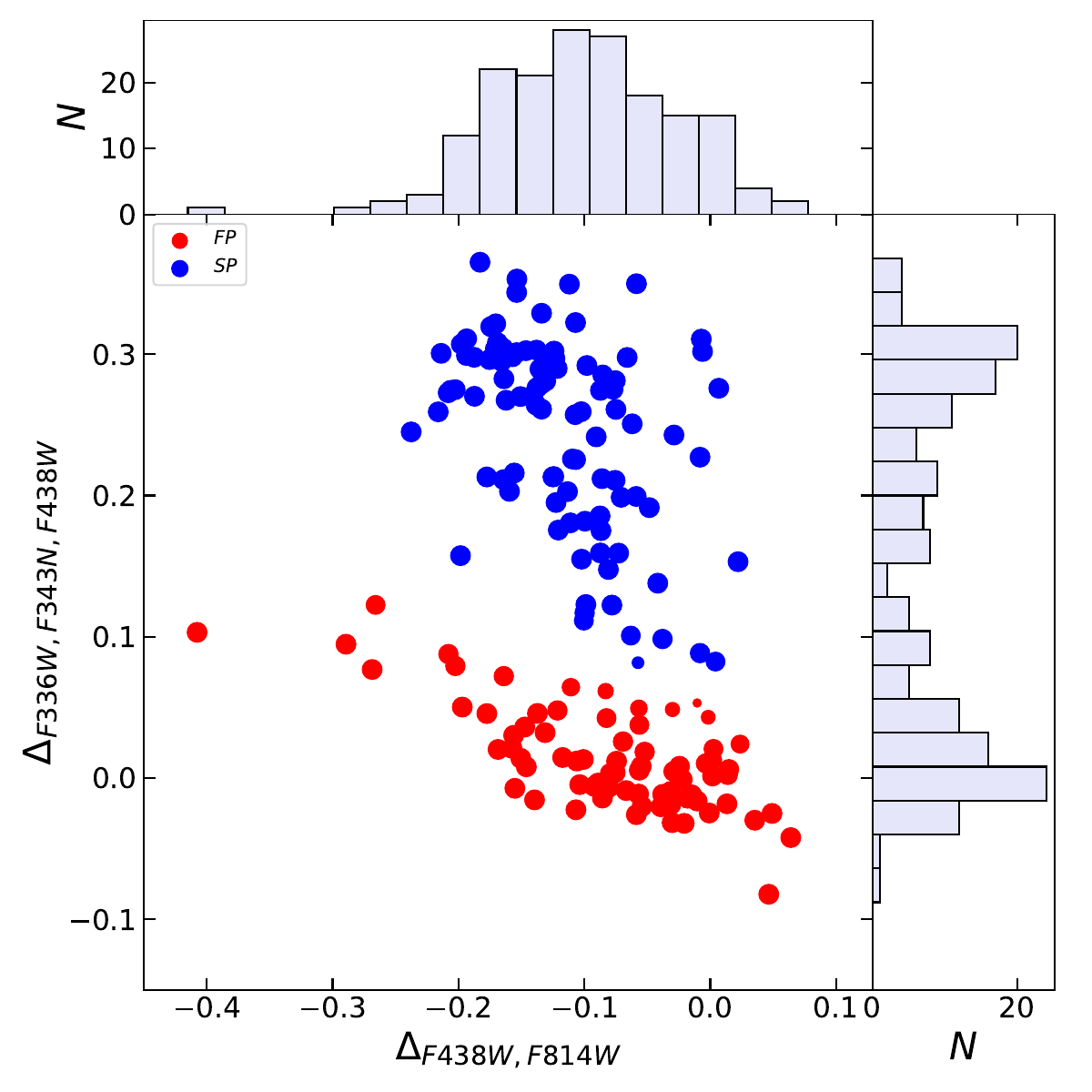}
\caption{ {\it Top panels:} ($m_{F814W}, m_{F275W}-m_{F814W}$) and ($m_{F814W}, C_{F275W,F336W,F438W}$) CMDs of NGC~3201. Data are from \citet{nardiello2018}. The red and blue lines in each panel represent the two fiducial lines at the edge of the RGB used to obtain the veriticalized color and pseudo-color distribution ($\Delta_{F275W,F814W}$ and $\Delta_{F275W,F336W,F438W}$). {\it Bottom panel:} The derived chromosome map. The upper and side panel show the histogram of ($\Delta_{F275W,F814W}$ and $\Delta_{F275W,F336W,F438W}$), respectively. Stars are colored in blue or red according to their classification as FP or SP stars, as resulted by the Gaussian mixture model fit. Stars are marked with probability dependent marker sizes: the larger is the probability that a star belong to the assigned population, the larger the marker size is.}
\label{fig:hstsel}
\end{figure}

We tagged FP and SP stars using an approach similar to that adopted in \citet[][see also \citealt{Milone2017}, \citealt{onorato23,cadelano23}]{Dalessandro2019}. Briefly, MPs were selected along the
RGB in the ($\Delta_{F275W,F814W}$, $\Delta_{F275W,F336W,F438W}$) diagram, the so-called “chromosome map”. We verticalized the distribution of RGB stars in the ($m_{F814W}$, $C_{F275W,F336W,F438W}$) and ($m_{F814W}, m_{F275W} - m_{F814W}$) diagrams (where $C_{F275W,F336W,F438W} =  (m_{F275W} - m_{F336W}) - (m_{F336W}-m_{F438W})$) with respect to two fiducial lines at the blue and red edges of the RGB in both CMDs (see left-hand Figure~\ref{fig:hstsel}). The combination of the two verticalized distributions gives the chromosome map presented in the right-hand panel of Figure~\ref{fig:hstsel}. The chromosome-map clearly reveals the presence of two distinct populations: the FP having low $\Delta_{F275W,F336W,F438W}$ and located across a large extension of $\Delta_{F275W,F814W}$ values, and the SP having high $\Delta_{F275W,F336W,F438W}$ values. The results are qualitatively in agreement with those obtained by \citet{Milone2017} and \citet{Kamann2020_n3201}. To refine the selection and assign to each star a probability of belonging to the FP or SP sub-populations, we fit the map with a 2D Gaussian mixture model using the \texttt{scikit-learn package}\footnote{\url{https://scikit-learn.org/stable/index.html}} \citep{scikit-learn}. The best-fit model is composed of two Gaussian functions which provide the separation shown with different colors in the Figure~\ref{fig:hstsel}.  {For each star, the sum of the probabilities of belonging to the FP and SP is equal to 1. Stars are tagged as FP/SP if their probability of belonging to the respective Gaussian is greater than 0.5. In this way,} 75 stars are assigned to the FP, while 97 are assigned to the SP.

\subsection{Wide-Field $C_{UBI}$ selection}

\begin{figure}[h!] 
\centering
\includegraphics[scale=0.28]{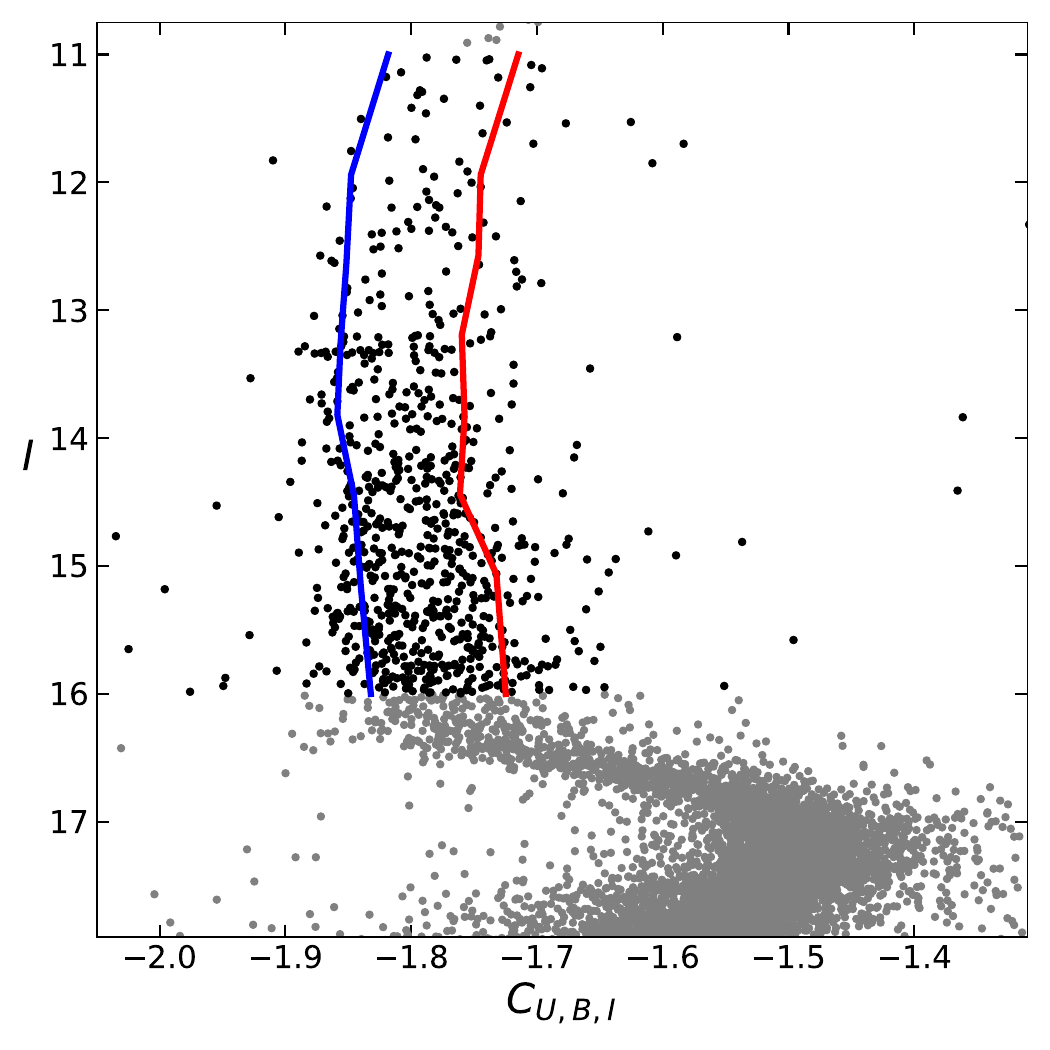}
\includegraphics[scale=0.28]{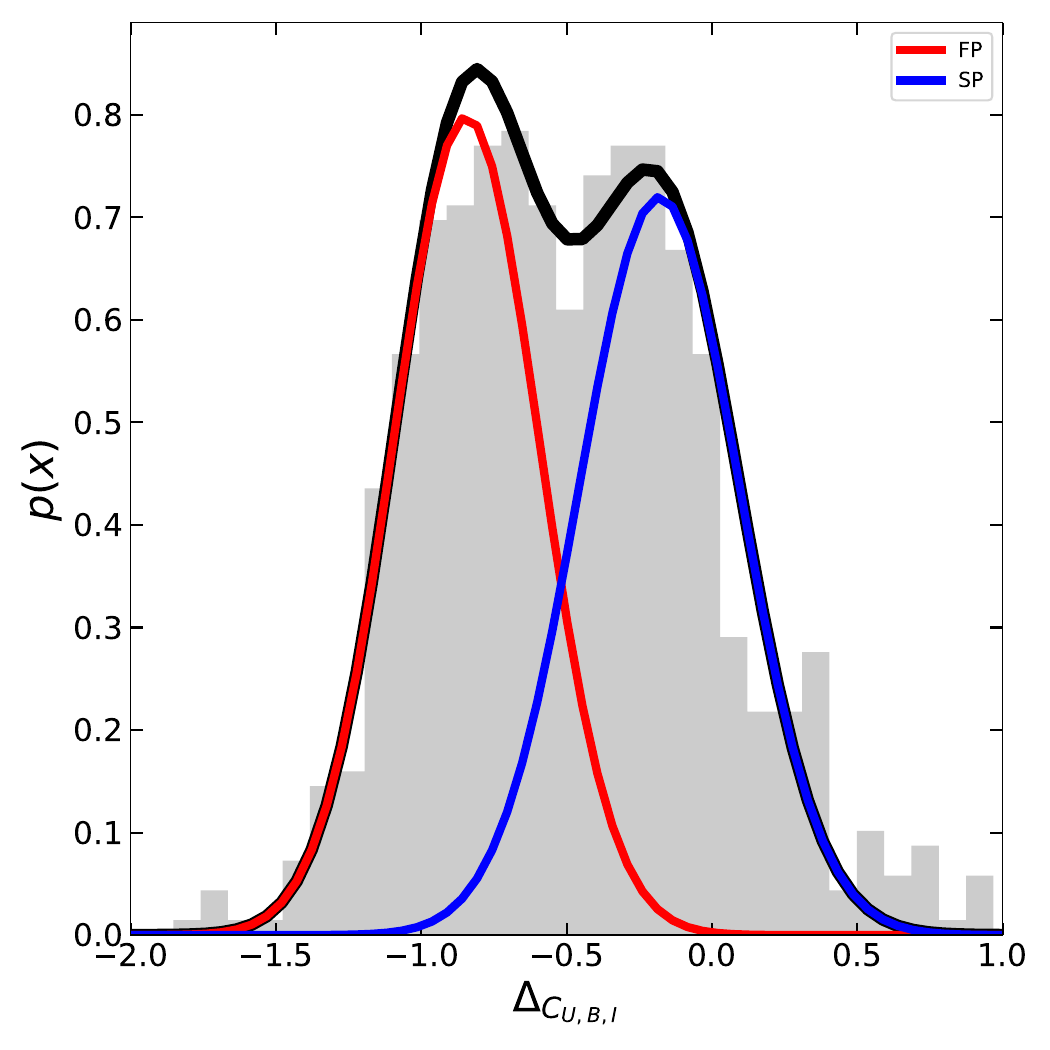}
\caption{{\it Top panel:} ($I$, $C_{U,B,I}$) CMD of NGC~3201. Data from \citet{stetson2019}. The red and blue lines are the two fiducial lines at the edge of the RGB used to obtain the verticalized $C_{U,B,I}$ distribution. {\it Bottom panel:} Histogram of the verticalized $C_{U,B,I}$ distribution. The red and blue curve are the two best-fit Gaussian used to select the FP and SP stars, respectively, while the black one is the sum of the two.}
\label{fig:cubisel}
\end{figure}

To separate the MPs in the wide-field catalog, RGB stars were verticalized in the  ($U$, $C_{U,B,I}$) CMD (where $C_{U,B,I}=(U-B)-(B-I)$) with respect to two fiducial lines on the blue and red edges of the sequence (see left-hand panel of Figure~\ref{fig:cubisel}). The distribution of the  $C_{U,B,I}$ pseudo-color is clearly bi-modal (see middle panel of Figure~\ref{fig:cubisel}). The two populations are separated around $C_{U,B,I}\approx-0.5$, where stars having lower and higher values are FP and SP stars, respectively. This distribution was fitted with a 2 component 1D Gaussian Mixture model to obtain the probability of each star to belong to the FP or SP. The best-fit Gaussian functions are shown in Figure~\ref{fig:cubisel}. 368 stars are assigned to the FP, while 406 are assigned to the SP. We used common stars between the HST and wide-field catalogue to ensure the consistency of the selection: the vast majority of FP (SP) stars as selected through the chromosome map are assigned to the corresponding population also in the $C_{U,B,I}$ selection. 

\section{Morphology of the Multiple Populations}
\begin{figure}[h] 
\centering
\includegraphics[scale=0.5]{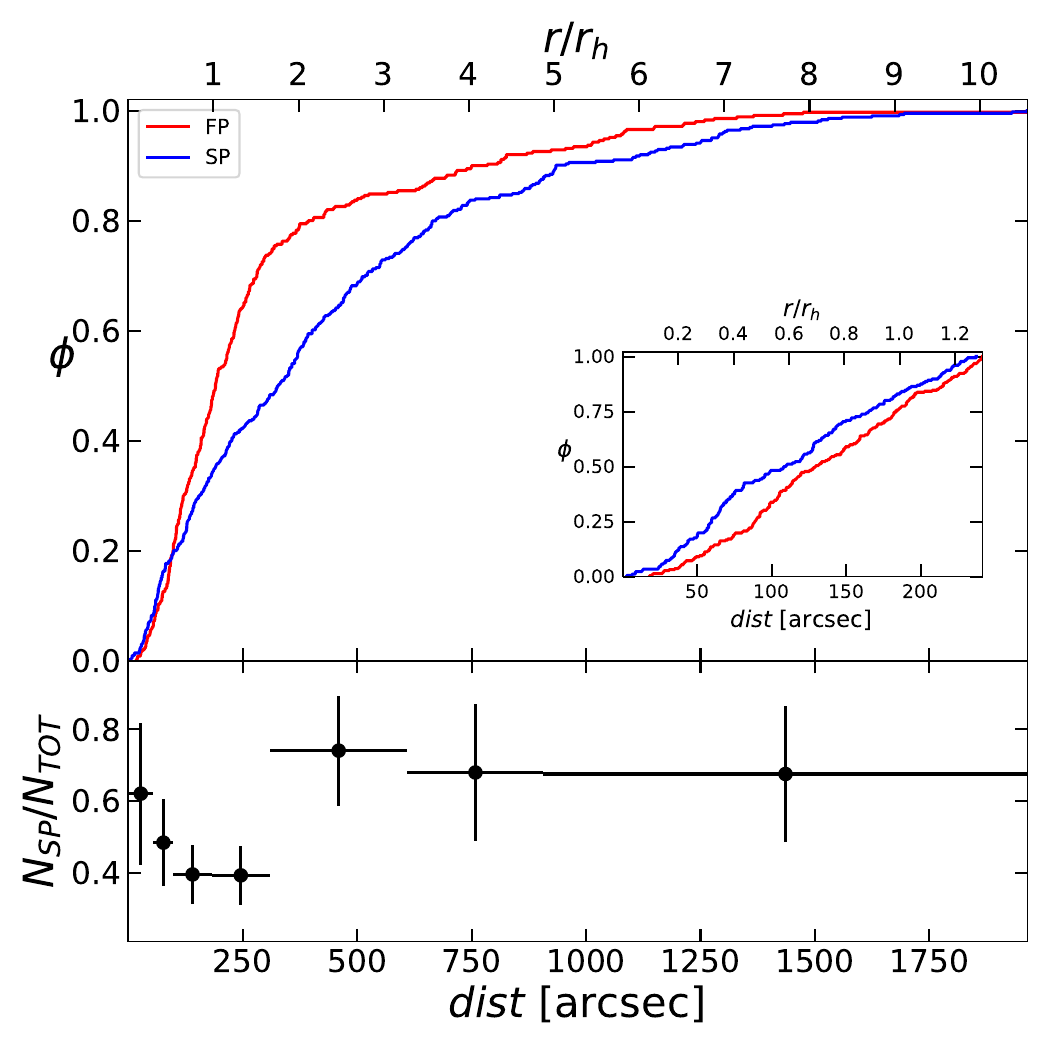}
\caption{{\it Top panel:}  Cumulative radial distribution of the FP (red curve) and SP (blue curve). The inset panel shows the cumulative distribution of stars within the cluster half-mass radius. {\it Bottom panel:} ratio between the number of SP stars $N_{SP}$ and the total amount of stars $N_{TOT}=N_{FP}+N_{SP}$ calculated in different radial bins.}
\label{fig:cumdistr}
\end{figure}

Figure~\ref{fig:cumdistr} shows the cumulative radial distribution of MPs. 
Results are in good agreement with those reported by \citet{leitinger2023} and this figure clearly shows why the value of $A^+$ calculated within $\sim4r_{h}$ (where $r_h$ is the half-mass radius $r_h=186\arcsec$, \citealp{ferraro2018}) is positive.
While a positive value of $A^+$ would in general correspond to a FP more centrally concentrated than the SP, the structural configuration of this cluster is actually more complex. 
However, the structural configuration is actually more complex.
As shown in the inset panel, within the half-mass radius we observe the opposite pattern: SP stars are more centrally concentrated than FP stars in the central region. 
This result is confirmed also by
the binned radial distribution of the ratio between the number of SPs (N$_{SP}$) to the total number of stars (N$_{TOT}$)  
as a function of cluster-centric distance as shown in the lower panel of Figure~\ref{fig:cumdistr}. 
In fact, such a distribution shows a puzzling bi-modal behavior, in agreement with the results previously reported by \citet{leitinger2023}.  We observe a central peak with $N_{SP}$/$N_{TOT}\sim0.6$, followed by a smooth decrease with a minimum at $\sim250\arcsec$ corresponding to $\sim1.3 r_h$  and an abrupt increase at distances larger than $\sim300\arcsec$ ($\sim1.6 r_h$), where $N_{SP}/N_{TOT}\sim0.7$.
It is important to stress here, that the central peak in the observed radial distribution is a prominent and significant feature that cannot be neglected in the overall analysis of the MP radial distributions. Indeed, it attains a significant fraction of the cluster extension, well behind its $r_h$, and it includes $\sim55\%$ of all the stars in the sample.  Moreover, according to the Kologorov-Smirov test, the probability that the two distributions within the half-mass radius  (inset in Figure~\ref{fig:cumdistr})  are extracted from the same parent distribution is of only $0.013\%$.

As discussed in the Introduction, since dynamical evolution is expected to have already partially smoothed out the primordial differences between FP and SP stars,
{\it the evidence that in the cluster central region SP stars are more centrally concentrated than the FP implies that at the time of cluster formation the SP must have been necessarily even more centrally segregated than observed today}. 

{Results are not significantly affected when a more conservative MP tagging, including only stars with a high probability of belonging to the assigned population, is adopted. }

\begin{figure*}[h] 
\centering
\includegraphics[scale=0.65]{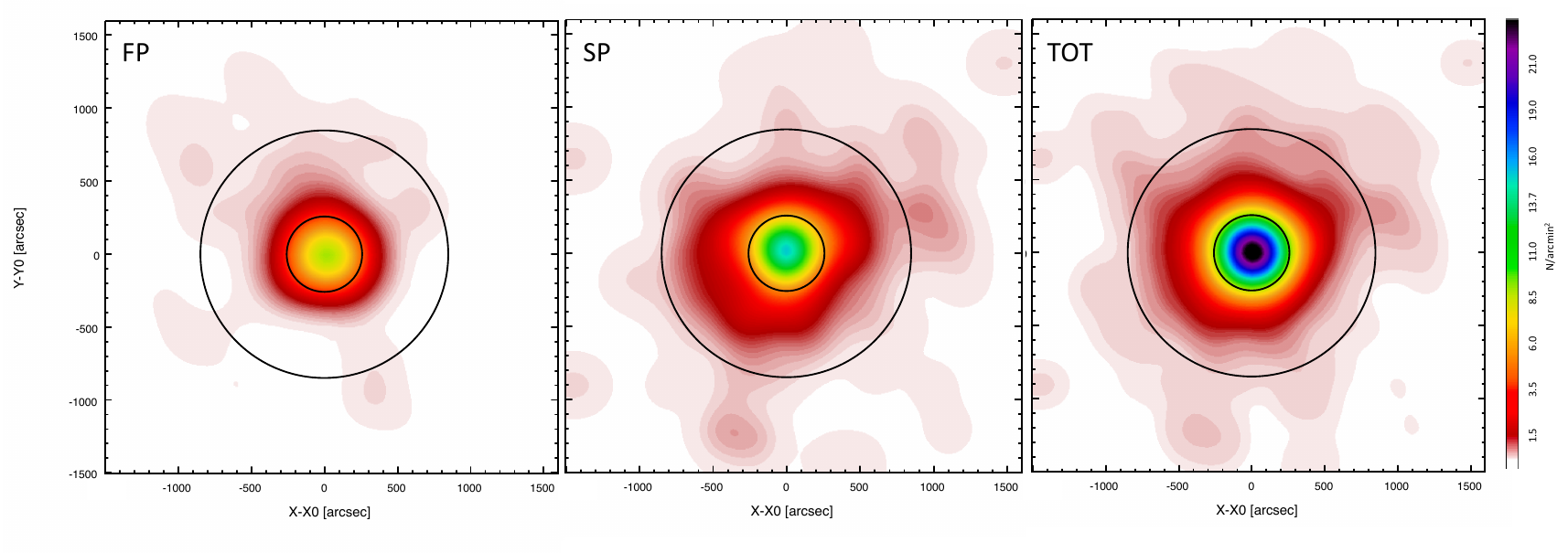}
\caption{Colour-coded surface density map of the analyzed field of view centered on the position of NGC~3201. The left panel shows the 2D map of the FP, the central panel that of SP stars and right panel the 2D map of total population (i.e. SP+FP). The inner and outer circles in each panel have a radius of $250\arcsec$ and $800\arcsec$, respectively. North is up, east is right.}
\label{fig:2dmap}
\end{figure*}

To delve deeper into this peculiar behavior, we analyzed the 2D surface density maps for MPs. To this aim, we transformed the distribution of selected RGB stars into a smoothed surface density function using a Gaussian kernel with a width of $75\arcsec$ in a grid covering the whole field of view and obtained with regular steps of $50\arcsec$ \citep[see, e.g.,][]{Dalessandro2015,onorato23,leanza23}. 
The 2D density map of the total sample of RGB stars shows an overall symmetrical/spherical morphology (right-hand panel of Figure~\ref{fig:2dmap}).
An overall symmetry is observed in the innermost regions ($<250\arcsec$) also for FP and SP stars (left-hand and middle panels in Figure~\ref{fig:2dmap}). Figure~\ref{fig:2dmap} also clearly shows that the SP population attains a more centrally concentrated distribution than the FP. 
FP and SP stars describe pretty asymmetrical and elongated distributions in the external regions. Such irregular distributions produce a pretty asymmetrical 2D $N_{SP}/N_{TOT}$ ratio distribution, which turns out to be particularly evident in in the radial range where $N_{SP}/N_{TOT}$ starts to increase again.
Such spatial variations can be so strong that the shape of the $N_{SP}/N_{TOT}$ 1D radial distribution can vary from being strongly bimodal to monotonically decreasing when only the N-W quadrant is considered. These asymmetries may arise from the intrinsic distribution of stars, the small number statistic in the outer cluster regions and uncorrected effects of differential reddening, or a combination of these factors. Regardless of their origin they cannot be overlooked in the overall analysis of the MP spatial distribution of this cluster.

\section{Kinematics of the Multiple Populations}

\begin{figure*}[h] 
\centering
\includegraphics[scale=0.3]{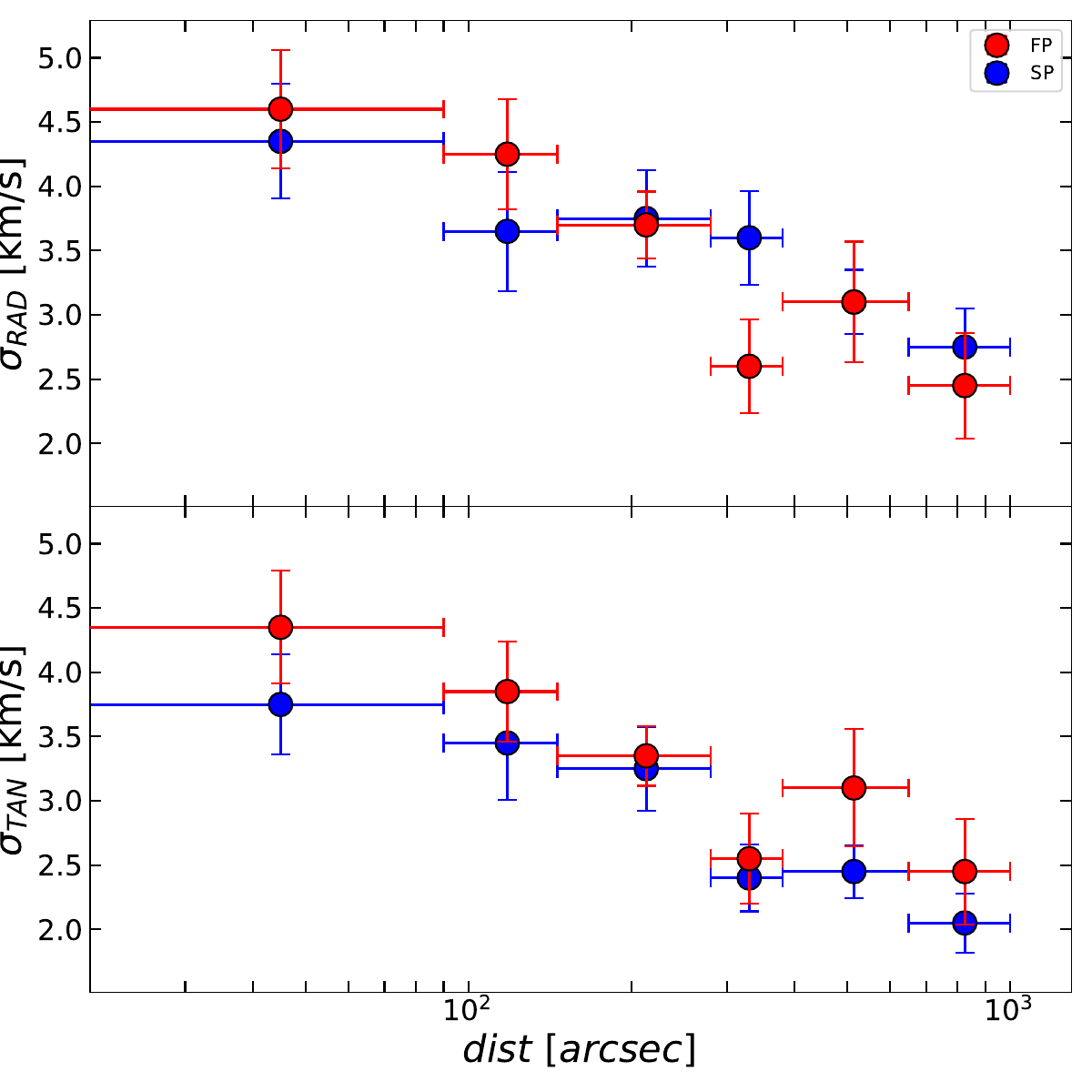}
\includegraphics[scale=0.3]{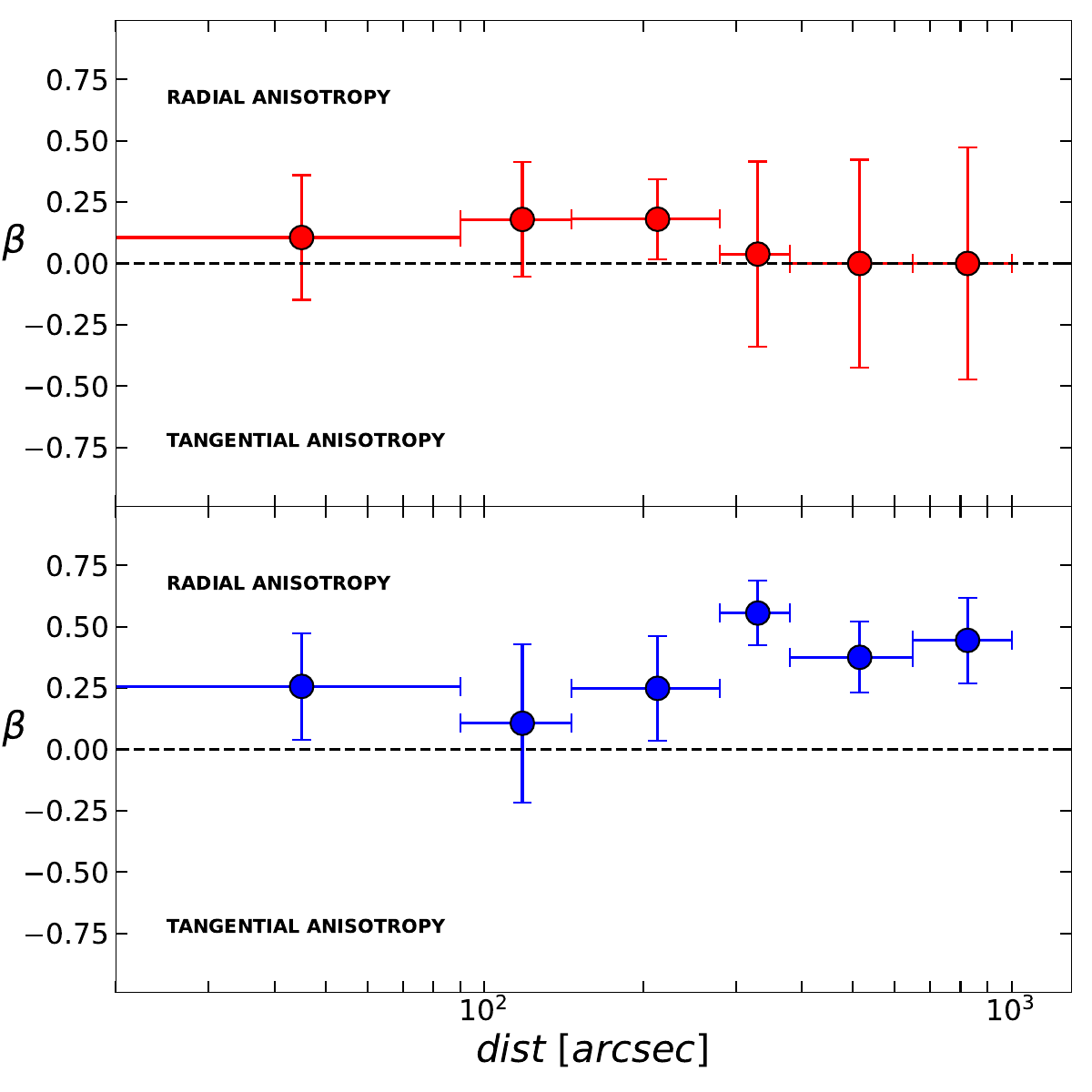}
\caption{Velocity dispersion and anisotropy profiles of the MPs in NGC~3201. {\it Left-hand panel:}   radial and tangential dispersion profiles are presented in the top and bottom panels, respectively. Red and blue points are obtained for FP and SP stars, respectively.  {\it Right-hand panel:} anisotropy profiles for the FP and SP stars in the top and bottom panel, respectively.}
\label{fig:vdisp}
\end{figure*}

The study of the kinematic properties of FP and SP stars can provide key insights into the formation and dynamical history of MPs. Indeed, as shown in a number of studies there is a close link between spatial and kinematic properties of FP and SP stars. Differences between the FP and the SP kinematics can be imprinted at at the time of SP formation (see e.g. \citealt{Bekki2010,Bekki2011,lacchin22}) or emerge during a cluster's evolution as a consequence of the initial differences between the FP and SP spatial distributions (see e.g. \citealt{Tiongco2019,Vesperini2021,Sollima2021}).

Here we analyze the kinematic of MPs by using the proper motions extracted from the Gaia DR3 catalogue. We selected stars with reliable proper motion measurements by following the selection criteria applied for the same cluster by \citet{bianchini2019}. In such a way, we have proper motions for 297 FP and 325 SP stars, which were converted to the tangential and radial components of the motion. 

To derive the velocity dispersion profiles of the populations along both the tangential and radial  components, we adopted the method fully described in \citet[][see also \citealt{raso20}]{Dalessandro2018a,dalessandro2021}. Briefly, the method consists in the measurement of the velocity dispersion for sample of stars within concentric radial bins and it is based on the maximum-likelihood estimator described by \citet{pryor1993}. It assumes that the probability of finding a star with a tangential/radial velocity $v_i \pm \epsilon_i$ at a projected distance from the cluster centre $r_i$ is described by a normal distribution:
\begin{equation}
    p(v_i,\epsilon_i,r_i)=\frac{1}{2\pi\sqrt{\sigma^2+\epsilon_i^2}}\exp \left[{-\frac{1}{2}\frac{(v_i-v_{sys})^2}{\sigma^2+\epsilon_i^2}}\right]
\end{equation}
where $v_{sys}$ and $\sigma$ are the systemic velocity and the intrinsic dispersion profile of the cluster, respectively. Following this approach, we measured the velocity dispersion in six concentric radial bins centered on the cluster centre and approximately containing the same number of stars. 

The resulting velocity dispersion profiles are shown in the left-hand panel of Figure~\ref{fig:vdisp}).
We then combined together the radial and tangential dispersion velocity measurements of each population to produce the anisotropy profile. The anisotropy was parameterized using $\beta=1-\sigma^2_{TAN}/\sigma^2_{RAD}$. Positive (negative) $\beta$ values indicate the occurrence of radial (tangential) anisotropy, while $\beta=0$ is consistent with an isotropic velocity distribution. The anisotropy profiles for FP and SP stars are presented in the top and bottom right-hand panel of Figure~\ref{fig:vdisp}, respectively. Interestingly, the two populations have quite a different $\beta$ profile. In fact, the FP profile shows an isotropic behavior, having $\beta$ values compatible with zero across the whole sampled field of view. On the contrary, SP stars are isotropic within the innermost $\sim200\arcsec$ from the center and then develop a clear radial anisotropy, reaching an average value of $\beta=0.46\pm0.07$ in the outer region. Such a behavior is exactly that expected from the diffusion of stars born more centrally concentrated that are progressively expanding on more radial orbits during the cluster dynamical evolution \citep{Bellini2015,Henault-Brunet2015,Vesperini2021,Sollima2021}.
Therefore, {\it the kinematic of SP stars in NGC~3201 provides further evidence that it was born more centrally concentrated than the FP of stars}, and would exclude that the opposite configuration was originally in place.

\section{Discussion and Summary}
All proposed scenarios for MP formation consistently predict the centrally concentrated formation of SP stars in the inner regions of a more spatially extended system of FP stars (see, e.g., \citealt{Decressin2007a,dercole08,Bekki2010,Bekki2011,Bastian2013,gieles18,lacchin22,Yaghoobi22a,Yaghoobi22b}). While the various dynamical processes acting during evolutionary phases following the formation epoch are expected to erase the initial dynamical distinctions between the populations, certain clusters might preserve some memory of these initial differences. Indeed, differences in the spatial and kinematic properties of FP and SP stars have been identified in several clusters (see, e.g., \citealt{Cordero2017,Simioni2016,Dalessandro2016,Dalessandro2018a,dalessandro2021,libralato23,onorato23}).
A comprehensive observational overview tracing the evolutionary path following the mixing of FP and SP stars was presented in the analysis by \citet{Dalessandro2019}. Their findings revealed that dynamically younger systems exhibit larger spatial differences between the populations, with the SP being more centrally concentrated than the FP, whereas dynamically evolved systems show spatially mixed MPs. Recently, \citet{leitinger2023} expanded this study to a broader cluster sample. They discovered that MPs in dynamically young systems can display any relative concentration, and notably, in at least two GCs (NGC~3201 and NGC~6101), the FP appears to be more centrally concentrated than the SP. Hence, it has been suggested that MPs can form with any possible initial relative concentration.

It is worth stressing that, as shown in several studies (see, e.g., \citealt{Dalessandro2019,Vesperini2021,Sollima2021}), even for dynamically young clusters, the present-day differences (or lack thereof) do not reflect the properties emerging at the end of the formation process. The mixing process is expected to start in the very early stages of a cluster's evolution, and the subsequent evolution over even just a few half-mass relaxation times further contributes to the mixing of MPs. Thus, while the present-day properties of a number of clusters may retain some memory of the primordial differences, in no case is the extent of these differences exactly that emerging at the end of the SP formation. As already pointed out in the Introduction, even small differences between the FP and SP spatial distributions are still consistent with scenarios predicting the SP formed more centrally concentrated than the FP.

In this paper, we directed our focus to NGC~3201, one of the two dynamically young clusters in the study of \citet{leitinger2023}. In this cluster, the SP appears to be less concentrated than the FP, posing a potential conflict with the predictions of all the MP formation scenarios proposed thus far. However, the results of our analysis suggest a more complex picture where the overall spatial and kinematic properties of MPs align with the SP forming more centrally concentrated than the FP. The conclusions drawn from our analysis are as follows.

\begin{itemize}
    \item The distribution of the SP is bimodal, as shown by the number ratio distribution over the global population  (see Figure~\ref{fig:cumdistr}). Within the cluster's central region and up to approximately $1.3r_h$, the SP is notably more centrally concentrated than the FP, with the SP fraction decreasing as distances from the cluster's center increase. This trend is in agreement with the predictions of multiple formation models, suggesting that SP stars initially formed in the inner regions of a more extended and less centrally concentrated FP system. Considering that dynamical processes have already started the mixing of the two populations, the present-day spatial differences are the remnants of stronger primordial ones. In the outer regions of the cluster ($r>2 r_{h}$), the SP fraction increases again (see Figure~\ref{fig:cumdistr}). Importantly, our analysis reveals that this outer increase of the SP fraction does not arise from a symmetric spatial distribution of FP and SP stars. Instead, it is likely due to a complex and irregular 2D distribution of the two populations.


\item We analyzed the cluster kinematics,  focusing specifically on the velocity anisotropy profile of the two populations. The FP turns out to be isotropic across all the sampled cluster extension. On the other hand, the SP is isotropic in the center and then develops radial anisotropy beyond the half-mass radius. This radial anisotropy is the expected kinematic signature of the SP's outward diffusion from an initially more centrally concentrated spatial distribution.  This additional kinematic feature provides further key evidence that the SP in the cluster was formed more centrally concentrated than FP.
\end{itemize}

The results derived in this study clearly show how a detailed analysis combining structural and kinematic observations allow key insights into the dynamical properties of MPs and are essential to constrain their possible formation and evolutionary paths. They also lend further support to scenarios in which the SP forms more centrally concentrated than the FP.

\begin{acknowledgements}
M.C. and E.D. acknowledge financial support from the project Light-on-Dark granted by MIUR through PRIN2017-2017K7REXT. E.D. acknowledges financial support from the Fulbright Visiting Scholar program 2023. E.D. is also grateful for the warm hospitality of the Indiana University where part of this work was performed.
E.V. acknowledges support from NSF grant AST-2009193. E.V. acknowledges also support from the John and A-Lan Reynolds Faculty Research Fund.
\end{acknowledgements}

%
\bibliographystyle{aa} 
\bibliography{ngc3201} 

\begin{thebibliography}{62}
\expandafter\ifx\csname natexlab\endcsname\relax\def\natexlab#1{#1}\fi

\bibitem[{{Bastian} {et~al.}(2013){Bastian}, {Lamers}, {de Mink}, {Longmore},
  {Goodwin}, \& {Gieles}}]{Bastian2013}
{Bastian}, N., {Lamers}, H.~J.~G.~L.~M., {de Mink}, S.~E., {et~al.} 2013,
  \mnras, 436, 2398

\bibitem[{{Bastian} \& {Lardo}(2018)}]{Bastian2018}
{Bastian}, N. \& {Lardo}, C. 2018, \araa, 56, 83

\bibitem[{{Bekki}(2010)}]{Bekki2010}
{Bekki}, K. 2010, \mnras, 401, L58

\bibitem[{{Bekki}(2011)}]{Bekki2011}
{Bekki}, K. 2011, \mnras, 412, 2241

\bibitem[{{Bellini} {et~al.}(2015){Bellini}, {Renzini}, {Anderson}, {Bedin},
  {Piotto}, {Soto}, {Brown}, {Milone}, {Sohn}, \& {Sweigart}}]{Bellini2015}
{Bellini}, A., {Renzini}, A., {Anderson}, J., {et~al.} 2015, \apj, 805, 178

\bibitem[{{Bianchini} {et~al.}(2019){Bianchini}, {Ibata}, \&
  {Famaey}}]{bianchini2019}
{Bianchini}, P., {Ibata}, R., \& {Famaey}, B. 2019, \apjl, 887, L12

\bibitem[{{Cadelano} {et~al.}(2022){Cadelano}, {Dalessandro}, {Salaris},
  {Bastian}, {Mucciarelli}, {Saracino}, {Martocchia}, \&
  {Cabrera-Ziri}}]{cadelano2022}
{Cadelano}, M., {Dalessandro}, E., {Salaris}, M., {et~al.} 2022, \apjl, 924, L2

\bibitem[{{Cadelano} {et~al.}(2020{\natexlab{a}}){Cadelano}, {Dalessandro},
  {Webb}, {Vesperini}, {Lattanzio}, {Beccari}, {Gomez}, \&
  {Monaco}}]{cadelano20b}
{Cadelano}, M., {Dalessandro}, E., {Webb}, J.~J., {et~al.} 2020{\natexlab{a}},
  \mnras, 499, 2390

\bibitem[{{Cadelano} {et~al.}(2023){Cadelano}, {Pallanca}, {Dalessandro},
  {Salaris}, {Mucciarelli}, {Leanza}, {Ferraro}, {Lanzoni}, {Chen}, {Freire},
  {Heinke}, \& {Ransom}}]{cadelano23}
{Cadelano}, M., {Pallanca}, C., {Dalessandro}, E., {et~al.} 2023, A\&A, 679, L13

\bibitem[{{Cadelano} {et~al.}(2020{\natexlab{b}}){Cadelano}, {Saracino},
  {Dalessandro}, {Ferraro}, {Lanzoni}, {Massari}, {Pallanca}, \&
  {Salaris}}]{cadelano2020_n6256}
{Cadelano}, M., {Saracino}, S., {Dalessandro}, E., {et~al.} 2020{\natexlab{b}},
  \apj, 895, 54

\bibitem[{{Calura} {et~al.}(2019){Calura}, {D'Ercole}, {Vesperini}, {Vanzella},
  \& {Sollima}}]{calura19}
{Calura}, F., {D'Ercole}, A., {Vesperini}, E., {Vanzella}, E., \& {Sollima}, A.
  2019, \mnras, 489, 3269

\bibitem[{{Cardelli} {et~al.}(1989){Cardelli}, {Clayton}, \&
  {Mathis}}]{cardelli89}
{Cardelli}, J.~A., {Clayton}, G.~C., \& {Mathis}, J.~S. 1989, \apj, 345, 245

\bibitem[{{Carretta} {et~al.}(2010){Carretta}, {Bragaglia}, {Gratton},
  {Lucatello}, {Bellazzini}, {Catanzaro}, {Leone}, {Momany}, {Piotto}, \&
  {D'Orazi}}]{Carretta2010}
{Carretta}, E., {Bragaglia}, A., {Gratton}, R.~G., {et~al.} 2010, \aap, 520,
  A95

\bibitem[{{Cordero} {et~al.}(2017){Cordero}, {H{\'e}nault-Brunet},
  {Pilachowski}, {Balbinot}, {Johnson}, \& {Varri}}]{Cordero2017}
{Cordero}, M.~J., {H{\'e}nault-Brunet}, V., {Pilachowski}, C.~A., {et~al.}
  2017, \mnras, 465, 3515

\bibitem[{{Dalessandro} {et~al.}(2019){Dalessandro}, {Cadelano}, {Vesperini},
  {Martocchia}, {Ferraro}, {Lanzoni}, {Bastian}, {Hong}, \&
  {Sanna}}]{Dalessandro2019}
{Dalessandro}, E., {Cadelano}, M., {Vesperini}, E., {et~al.} 2019, \apjl, 884,
  L24

\bibitem[{{Dalessandro} {et~al.}(2018{\natexlab{a}}){Dalessandro}, {Cadelano},
  {Vesperini}, {Salaris}, {Ferraro}, {Lanzoni}, {Raso}, {Hong}, {Webb}, \&
  {Zocchi}}]{Dalessandro2018a}
{Dalessandro}, E., {Cadelano}, M., {Vesperini}, E., {et~al.}
  2018{\natexlab{a}}, \apj, 859, 15

\bibitem[{{Dalessandro} {et~al.}(2016){Dalessandro}, {Lapenna}, {Mucciarelli},
  {Origlia}, {Ferraro}, \& {Lanzoni}}]{Dalessandro2016}
{Dalessandro}, E., {Lapenna}, E., {Mucciarelli}, A., {et~al.} 2016, \apj, 829,
  77

\bibitem[{{Dalessandro} {et~al.}(2018{\natexlab{b}}){Dalessandro}, {Lardo},
  {Cadelano}, {Saracino}, {Bastian}, {Mucciarelli}, {Salaris}, {Stetson}, \&
  {Pancino}}]{dalessandro2018b}
{Dalessandro}, E., {Lardo}, C., {Cadelano}, M., {et~al.} 2018{\natexlab{b}},
  \aap, 618, A131

\bibitem[{{Dalessandro} {et~al.}(2015){Dalessandro}, {Miocchi}, {Carraro},
  {J{\'\i}lkov{\'a}}, \& {Moitinho}}]{Dalessandro2015}
{Dalessandro}, E., {Miocchi}, P., {Carraro}, G., {J{\'\i}lkov{\'a}}, L., \&
  {Moitinho}, A. 2015, \mnras, 449, 1811

\bibitem[{{Dalessandro} {et~al.}(2018{\natexlab{c}}){Dalessandro},
  {Mucciarelli}, {Bellazzini}, {Sollima}, {Vesperini}, {Hong},
  {H{\'e}nault-Brunet}, {Ferraro}, {Ibata}, {Lanzoni}, {Massari}, \&
  {Salaris}}]{Dalessandro2018c}
{Dalessandro}, E., {Mucciarelli}, A., {Bellazzini}, M., {et~al.}
  2018{\natexlab{c}}, \apj, 864, 33

\bibitem[{{Dalessandro} {et~al.}(2021){Dalessandro}, {Raso}, {Kamann},
  {Bellazzini}, {Vesperini}, {Bellini}, \& {Beccari}}]{dalessandro2021}
{Dalessandro}, E., {Raso}, S., {Kamann}, S., {et~al.} 2021, \mnras, 506, 813

\bibitem[{{D'Antona} {et~al.}(2016){D'Antona}, {Vesperini}, {D'Ercole},
  {Ventura}, {Milone}, {Marino}, \& {Tailo}}]{dantona16}
{D'Antona}, F., {Vesperini}, E., {D'Ercole}, A., {et~al.} 2016, \mnras, 458,
  2122

\bibitem[{{de Mink} {et~al.}(2009){de Mink}, {Pols}, {Langer}, \&
  {Izzard}}]{demink09}
{de Mink}, S.~E., {Pols}, O.~R., {Langer}, N., \& {Izzard}, R.~G. 2009, \aap,
  507, L1

\bibitem[{{Decressin} {et~al.}(2007){Decressin}, {Charbonnel}, \&
  {Meynet}}]{Decressin2007a}
{Decressin}, T., {Charbonnel}, C., \& {Meynet}, G. 2007, \aap, 475, 859

\bibitem[{{Denissenkov} \& {Hartwick}(2014)}]{Denissenkov2014}
{Denissenkov}, P.~A. \& {Hartwick}, F.~D.~A. 2014, \mnras, 437, L21

\bibitem[{{Deras} {et~al.}(2023){Deras}, {Cadelano}, {Ferraro}, {Lanzoni}, \&
  {Pallanca}}]{deras2023}
{Deras}, D., {Cadelano}, M., {Ferraro}, F.~R., {Lanzoni}, B., \& {Pallanca}, C.
  2023, \apj, 942, 104

\bibitem[{{D'Ercole} {et~al.}(2008){D'Ercole}, {Vesperini}, {D'Antona},
  {McMillan}, \& {Recchi}}]{dercole08}
{D'Ercole}, A., {Vesperini}, E., {D'Antona}, F., {McMillan}, S. L.~W., \&
  {Recchi}, S. 2008, \mnras, 391, 825

\bibitem[{{Ferraro} {et~al.}(2018){Ferraro}, {Mucciarelli}, {Lanzoni},
  {Pallanca}, {Lapenna}, {Origlia}, {Dalessandro}, {Valenti}, {Beccari},
  {Bellazzini}, {Vesperini}, {Varri}, \& {Sollima}}]{ferraro2018}
{Ferraro}, F.~R., {Mucciarelli}, A., {Lanzoni}, B., {et~al.} 2018, \apj, 860,
  50

\bibitem[{{Gaia Collaboration} {et~al.}(2023){Gaia Collaboration}, {Vallenari},
  {Brown}, {Prusti}, {de Bruijne}, {Arenou}, {Babusiaux}, {Biermann},
  {Creevey}, {Ducourant}, {Evans}, {Eyer}, {Guerra}, {Hutton}, {Jordi},
  {Klioner}, {Lammers}, {Lindegren}, {Luri}, {Mignard}, {Panem}, {Pourbaix},
  {Randich}, {Sartoretti}, {Soubiran}, {Tanga}, {Walton}, {Bailer-Jones},
  {Bastian}, {Drimmel}, {Jansen}, {Katz}, {Lattanzi}, {van Leeuwen}, {Bakker},
  {Cacciari}, {Casta{\~n}eda}, {De Angeli}, {Fabricius}, {Fouesneau},
  {Fr{\'e}mat}, {Galluccio}, {Guerrier}, {Heiter}, {Masana}, {Messineo},
  {Mowlavi}, {Nicolas}, {Nienartowicz}, {Pailler}, {Panuzzo}, {Riclet}, {Roux},
  {Seabroke}, {Sordo}, {Th{\'e}venin}, {Gracia-Abril}, {Portell}, {Teyssier},
  {Altmann}, {Andrae}, {Audard}, {Bellas-Velidis}, {Benson}, {Berthier},
  {Blomme}, {Burgess}, {Busonero}, {Busso}, {C{\'a}novas}, {Carry}, {Cellino},
  {Cheek}, {Clementini}, {Damerdji}, {Davidson}, {de Teodoro}, {Nu{\~n}ez
  Campos}, {Delchambre}, {Dell'Oro}, {Esquej}, {Fern{\'a}ndez-Hern{\'a}ndez},
  {Fraile}, {Garabato}, {Garc{\'\i}a-Lario}, {Gosset}, {Haigron}, {Halbwachs},
  {Hambly}, {Harrison}, {Hern{\'a}ndez}, {Hestroffer}, {Hodgkin}, {Holl},
  {Jan{\ss}en}, {Jevardat de Fombelle}, {Jordan}, {Krone-Martins}, {Lanzafame},
  {L{\"o}ffler}, {Marchal}, {Marrese}, {Moitinho}, {Muinonen}, {Osborne},
  {Pancino}, {Pauwels}, {Recio-Blanco}, {Reyl{\'e}}, {Riello}, {Rimoldini},
  {Roegiers}, {Rybizki}, {Sarro}, {Siopis}, {Smith}, {Sozzetti}, {Utrilla},
  {van Leeuwen}, {Abbas}, {{\'A}brah{\'a}m}, {Abreu Aramburu}, {Aerts},
  {Aguado}, {Ajaj}, {Aldea-Montero}, {Altavilla}, {{\'A}lvarez}, {Alves},
  {Anders}, {Anderson}, {Anglada Varela}, {Antoja}, {Baines}, {Baker},
  {Balaguer-N{\'u}{\~n}ez}, {Balbinot}, {Balog}, {Barache}, {Barbato},
  {Barros}, {Barstow}, {Bartolom{\'e}}, {Bassilana}, {Bauchet}, {Becciani},
  {Bellazzini}, {Berihuete}, {Bernet}, {Bertone}, {Bianchi}, {Binnenfeld},
  {Blanco-Cuaresma}, {Blazere}, {Boch}, {Bombrun}, {Bossini}, {Bouquillon},
  {Bragaglia}, {Bramante}, {Breedt}, {Bressan}, {Brouillet}, {Brugaletta},
  {Bucciarelli}, {Burlacu}, {Butkevich}, {Buzzi}, {Caffau}, {Cancelliere},
  {Cantat-Gaudin}, {Carballo}, {Carlucci}, {Carnerero}, {Carrasco},
  {Casamiquela}, {Castellani}, {Castro-Ginard}, {Chaoul}, {Charlot}, {Chemin},
  {Chiaramida}, {Chiavassa}, {Chornay}, {Comoretto}, {Contursi}, {Cooper},
  {Cornez}, {Cowell}, {Crifo}, {Cropper}, {Crosta}, {Crowley}, {Dafonte},
  {Dapergolas}, {David}, {David}, {de Laverny}, {De Luise}, {De March}, {De
  Ridder}, {de Souza}, {de Torres}, {del Peloso}, {del Pozo}, {Delbo},
  {Delgado}, {Delisle}, {Demouchy}, {Dharmawardena}, {Di Matteo}, {Diakite},
  {Diener}, {Distefano}, {Dolding}, {Edvardsson}, {Enke}, {Fabre}, {Fabrizio},
  {Faigler}, {Fedorets}, {Fernique}, {Fienga}, {Figueras}, {Fournier},
  {Fouron}, {Fragkoudi}, {Gai}, {Garcia-Gutierrez}, {Garcia-Reinaldos},
  {Garc{\'\i}a-Torres}, {Garofalo}, {Gavel}, {Gavras}, {Gerlach}, {Geyer},
  {Giacobbe}, {Gilmore}, {Girona}, {Giuffrida}, {Gomel}, {Gomez},
  {Gonz{\'a}lez-N{\'u}{\~n}ez}, {Gonz{\'a}lez-Santamar{\'\i}a},
  {Gonz{\'a}lez-Vidal}, {Granvik}, {Guillout}, {Guiraud},
  {Guti{\'e}rrez-S{\'a}nchez}, {Guy}, {Hatzidimitriou}, {Hauser}, {Haywood},
  {Helmer}, {Helmi}, {Sarmiento}, {Hidalgo}, {Hilger}, {H{\l}adczuk}, {Hobbs},
  {Holland}, {Huckle}, {Jardine}, {Jasniewicz}, {Jean-Antoine Piccolo},
  {Jim{\'e}nez-Arranz}, {Jorissen}, {Juaristi Campillo}, {Julbe}, {Karbevska},
  {Kervella}, {Khanna}, {Kontizas}, {Kordopatis}, {Korn}, {K{\'o}sp{\'a}l},
  {Kostrzewa-Rutkowska}, {Kruszy{\'n}ska}, {Kun}, {Laizeau}, {Lambert},
  {Lanza}, {Lasne}, {Le Campion}, {Lebreton}, {Lebzelter}, {Leccia}, {Leclerc},
  {Lecoeur-Taibi}, {Liao}, {Licata}, {Lindstr{\o}m}, {Lister}, {Livanou},
  {Lobel}, {Lorca}, {Loup}, {Madrero Pardo}, {Magdaleno Romeo}, {Managau},
  {Mann}, {Manteiga}, {Marchant}, {Marconi}, {Marcos}, {Marcos Santos},
  {Mar{\'\i}n Pina}, {Marinoni}, {Marocco}, {Marshall}, {Martin Polo},
  {Mart{\'\i}n-Fleitas}, {Marton}, {Mary}, {Masip}, {Massari},
  {Mastrobuono-Battisti}, {Mazeh}, {McMillan}, {Messina}, {Michalik}, {Millar},
  {Mints}, {Molina}, {Molinaro}, {Moln{\'a}r}, {Monari}, {Mongui{\'o}},
  {Montegriffo}, {Montero}, {Mor}, {Mora}, {Morbidelli}, {Morel}, {Morris},
  {Muraveva}, {Murphy}, {Musella}, {Nagy}, {Noval}, {Oca{\~n}a}, {Ogden},
  {Ordenovic}, {Osinde}, {Pagani}, {Pagano}, {Palaversa}, {Palicio},
  {Pallas-Quintela}, {Panahi}, {Payne-Wardenaar}, {Pe{\~n}alosa Esteller},
  {Penttil{\"a}}, {Pichon}, {Piersimoni}, {Pineau}, {Plachy}, {Plum}, {Poggio},
  {Pr{\v{s}}a}, {Pulone}, {Racero}, {Ragaini}, {Rainer}, {Raiteri}, {Rambaux},
  {Ramos}, {Ramos-Lerate}, {Re Fiorentin}, {Regibo}, {Richards}, {Rios Diaz},
  {Ripepi}, {Riva}, {Rix}, {Rixon}, {Robichon}, {Robin}, {Robin}, {Roelens},
  {Rogues}, {Rohrbasser}, {Romero-G{\'o}mez}, {Rowell}, {Royer}, {Ruz Mieres},
  {Rybicki}, {Sadowski}, {S{\'a}ez N{\'u}{\~n}ez}, {Sagrist{\`a} Sell{\'e}s},
  {Sahlmann}, {Salguero}, {Samaras}, {Sanchez Gimenez}, {Sanna},
  {Santove{\~n}a}, {Sarasso}, {Schultheis}, {Sciacca}, {Segol}, {Segovia},
  {S{\'e}gransan}, {Semeux}, {Shahaf}, {Siddiqui}, {Siebert}, {Siltala},
  {Silvelo}, {Slezak}, {Slezak}, {Smart}, {Snaith}, {Solano}, {Solitro},
  {Souami}, {Souchay}, {Spagna}, {Spina}, {Spoto}, {Steele},
  {Steidelm{\"u}ller}, {Stephenson}, {S{\"u}veges}, {Surdej}, {Szabados},
  {Szegedi-Elek}, {Taris}, {Taylor}, {Teixeira}, {Tolomei}, {Tonello}, {Torra},
  {Torra}, {Torralba Elipe}, {Trabucchi}, {Tsounis}, {Turon}, {Ulla}, {Unger},
  {Vaillant}, {van Dillen}, {van Reeven}, {Vanel}, {Vecchiato}, {Viala},
  {Vicente}, {Voutsinas}, {Weiler}, {Wevers}, {Wyrzykowski}, {Yoldas}, {Yvard},
  {Zhao}, {Zorec}, {Zucker}, \& {Zwitter}}]{gaia_dr3}
{Gaia Collaboration}, {Vallenari}, A., {Brown}, A.~G.~A., {et~al.} 2023, \aap,
  674, A1

\bibitem[{{Gieles} {et~al.}(2018){Gieles}, {Charbonnel}, {Krause},
  {H{\'e}nault-Brunet}, {Agertz}, {Lamers}, {Bastian}, {Gualandris}, {Zocchi},
  \& {Petts}}]{gieles18}
{Gieles}, M., {Charbonnel}, C., {Krause}, M. G.~H., {et~al.} 2018, \mnras, 478,
  2461

\bibitem[{{Girardi} {et~al.}(2002){Girardi}, {Bertelli}, {Bressan}, {Chiosi},
  {Groenewegen}, {Marigo}, {Salasnich}, \& {Weiss}}]{girardi02}
{Girardi}, L., {Bertelli}, G., {Bressan}, A., {et~al.} 2002, \aap, 391, 195

\bibitem[{{Goldsbury} {et~al.}(2010){Goldsbury}, {Richer}, {Anderson},
  {Dotter}, {Sarajedini}, \& {Woodley}}]{goldsbury10}
{Goldsbury}, R., {Richer}, H.~B., {Anderson}, J., {et~al.} 2010, \aj, 140, 1830

\bibitem[{{Gratton} {et~al.}(2019){Gratton}, {Bragaglia}, {Carretta},
  {D'Orazi}, {Lucatello}, \& {Sollima}}]{gratton19}
{Gratton}, R., {Bragaglia}, A., {Carretta}, E., {et~al.} 2019, \aapr, 27, 8

\bibitem[{{H{\'e}nault-Brunet} {et~al.}(2015){H{\'e}nault-Brunet}, {Gieles},
  {Agertz}, \& {Read}}]{Henault-Brunet2015}
{H{\'e}nault-Brunet}, V., {Gieles}, M., {Agertz}, O., \& {Read}, J.~I. 2015,
  \mnras, 450, 1164

\bibitem[{{Kamann} {et~al.}(2020){Kamann}, {Giesers}, {Bastian}, {Brinchmann},
  {Dreizler}, {G{\"o}ttgens}, {Husser}, {Latour}, {Weilbacher}, \&
  {Wisotzki}}]{Kamann2020_n3201}
{Kamann}, S., {Giesers}, B., {Bastian}, N., {et~al.} 2020, \aap, 635, A65

\bibitem[{{Lacchin} {et~al.}(2022){Lacchin}, {Calura}, {Vesperini}, \&
  {Mastrobuono-Battisti}}]{lacchin22}
{Lacchin}, E., {Calura}, F., {Vesperini}, E., \& {Mastrobuono-Battisti}, A.
  2022, \mnras, 517, 1171

\bibitem[{{Larsen} {et~al.}(2014){Larsen}, {Brodie}, {Grundahl}, \&
  {Strader}}]{larsen14}
{Larsen}, S.~S., {Brodie}, J.~P., {Grundahl}, F., \& {Strader}, J. 2014, \apj,
  797, 15

\bibitem[{{Leanza} {et~al.}(2023){Leanza}, {Pallanca}, {Ferraro}, {Lanzoni},
  {Dalessandro}, {Cadelano}, {Vesperini}, {Origlia}, {Mucciarelli}, \&
  {Valenti}}]{leanza23}
{Leanza}, S., {Pallanca}, C., {Ferraro}, F.~R., {et~al.} 2023, \apj, 944, 162

\bibitem[{{Leitinger} {et~al.}(2023){Leitinger}, {Baumgardt}, {Cabrera-Ziri},
  {Hilker}, \& {Pancino}}]{leitinger2023}
{Leitinger}, E., {Baumgardt}, H., {Cabrera-Ziri}, I., {Hilker}, M., \&
  {Pancino}, E. 2023, \mnras, 520, 1456

\bibitem[{{Libralato} {et~al.}(2023){Libralato}, {Vesperini}, {Bellini},
  {Milone}, {van der Marel}, {Piotto}, {Anderson}, {Aparicio}, {Barbuy},
  {Bedin}, {Brown}, {Cassisi}, {Nardiello}, {Sarajedini}, \&
  {Scalco}}]{libralato23}
{Libralato}, M., {Vesperini}, E., {Bellini}, A., {et~al.} 2023, \apj, 944, 58

\bibitem[{{Martocchia} {et~al.}(2018){Martocchia}, {Niederhofer},
  {Dalessandro}, {Bastian}, {Kacharov}, {Usher}, {Cabrera-Ziri}, {Lardo},
  {Cassisi}, {Geisler}, {Hilker}, {Hollyhead}, {Kozhurina-Platais}, {Larsen},
  {Mackey}, {Mucciarelli}, {Platais}, \& {Salaris}}]{Martocchia2018}
{Martocchia}, S., {Niederhofer}, F., {Dalessandro}, E., {et~al.} 2018, \mnras,
  477, 4696

\bibitem[{{Miholics} {et~al.}(2015){Miholics}, {Webb}, \& {Sills}}]{miholics15}
{Miholics}, M., {Webb}, J.~J., \& {Sills}, A. 2015, \mnras, 454, 2166

\bibitem[{{Milone} {et~al.}(2017){Milone}, {Piotto}, {Renzini}, {Marino},
  {Bedin}, {Vesperini}, {D'Antona}, {Nardiello}, {Anderson}, {King}, {Yong},
  {Bellini}, {Aparicio}, {Barbuy}, {Brown}, {Cassisi}, {Ortolani}, {Salaris},
  {Sarajedini}, \& {van der Marel}}]{Milone2017}
{Milone}, A.~P., {Piotto}, G., {Renzini}, A., {et~al.} 2017, \mnras, 464, 3636

\bibitem[{{Mucciarelli} {et~al.}(2008){Mucciarelli}, {Carretta}, {Origlia}, \&
  {Ferraro}}]{mucciarelli08}
{Mucciarelli}, A., {Carretta}, E., {Origlia}, L., \& {Ferraro}, F.~R. 2008,
  \aj, 136, 375

\bibitem[{{Nardiello} {et~al.}(2018){Nardiello}, {Libralato}, {Piotto},
  {Anderson}, {Bellini}, {Aparicio}, {Bedin}, {Cassisi}, {Granata}, {King},
  {Lucertini}, {Marino}, {Milone}, {Ortolani}, {Platais}, \& {van der
  Marel}}]{nardiello2018}
{Nardiello}, D., {Libralato}, M., {Piotto}, G., {et~al.} 2018, \mnras, 481,
  3382

\bibitem[{{Onorato} {et~al.}(2023){Onorato}, {Cadelano}, {Dalessandro},
  {Vesperini}, {Lanzoni}, \& {Mucciarelli}}]{onorato23}
{Onorato}, S., {Cadelano}, M., {Dalessandro}, E., {et~al.} 2023, \aap, 677, A8

\bibitem[{Pedregosa {et~al.}(2011)Pedregosa, Varoquaux, Gramfort, Michel,
  Thirion, Grisel, Blondel, Prettenhofer, Weiss, Dubourg, Vanderplas, Passos,
  Cournapeau, Brucher, Perrot, \& Duchesnay}]{scikit-learn}
Pedregosa, F., Varoquaux, G., Gramfort, A., {et~al.} 2011, Journal of Machine
  Learning Research, 12, 2825

\bibitem[{{Piotto} {et~al.}(2015){Piotto}, {Milone}, {Bedin}, {Anderson},
  {King}, {Marino}, {Nardiello}, {Aparicio}, {Barbuy}, {Bellini}, {Brown},
  {Cassisi}, {Cool}, {Cunial}, {Dalessandro}, {D'Antona}, {Ferraro}, {Hidalgo},
  {Lanzoni}, {Monelli}, {Ortolani}, {Renzini}, {Salaris}, {Sarajedini}, {van
  der Marel}, {Vesperini}, \& {Zoccali}}]{piotto2015}
{Piotto}, G., {Milone}, A.~P., {Bedin}, L.~R., {et~al.} 2015, \aj, 149, 91

\bibitem[{{Pryor} \& {Meylan}(1993)}]{pryor1993}
{Pryor}, C. \& {Meylan}, G. 1993, in Astronomical Society of the Pacific
  Conference Series, Vol.~50, Structure and Dynamics of Globular Clusters, ed.
  S.~G. {Djorgovski} \& G.~{Meylan}, 357

\bibitem[{{Raso} {et~al.}(2020){Raso}, {Libralato}, {Bellini}, {Ferraro},
  {Lanzoni}, {Cadelano}, {Pallanca}, {Dalessandro}, {Piotto}, {Anderson}, \&
  {Sohn}}]{raso20}
{Raso}, S., {Libralato}, M., {Bellini}, A., {et~al.} 2020, \apj, 895, 15

\bibitem[{{Renzini} {et~al.}(2022){Renzini}, {Marino}, \& {Milone}}]{renzini22}
{Renzini}, A., {Marino}, A.~F., \& {Milone}, A.~P. 2022, \mnras, 513, 2111

\bibitem[{{Sills} {et~al.}(2019){Sills}, {Dalessandro}, {Cadelano},
  {Alfaro-Cuello}, \& {Kruijssen}}]{sills19}
{Sills}, A., {Dalessandro}, E., {Cadelano}, M., {Alfaro-Cuello}, M., \&
  {Kruijssen}, J.~M.~D. 2019, \mnras, 490, L67

\bibitem[{{Simioni} {et~al.}(2016){Simioni}, {Milone}, {Bedin}, {Aparicio},
  {Piotto}, {Vesperini}, \& {Hong}}]{Simioni2016}
{Simioni}, M., {Milone}, A.~P., {Bedin}, L.~R., {et~al.} 2016, \mnras, 463, 449

\bibitem[{{Sollima}(2021)}]{Sollima2021}
{Sollima}, A. 2021, \mnras, 502, 1974

\bibitem[{{Stetson} {et~al.}(2019){Stetson}, {Pancino}, {Zocchi}, {Sanna}, \&
  {Monelli}}]{stetson2019}
{Stetson}, P.~B., {Pancino}, E., {Zocchi}, A., {Sanna}, N., \& {Monelli}, M.
  2019, \mnras, 485, 3042

\bibitem[{{Tiongco} {et~al.}(2019){Tiongco}, {Vesperini}, \&
  {Varri}}]{Tiongco2019}
{Tiongco}, M.~A., {Vesperini}, E., \& {Varri}, A.~L. 2019, \mnras, 487, 5535

\bibitem[{{Vesperini} {et~al.}(2021){Vesperini}, {Hong}, {Giersz}, \&
  {Hypki}}]{Vesperini2021}
{Vesperini}, E., {Hong}, J., {Giersz}, M., \& {Hypki}, A. 2021, \mnras, 502,
  4290

\bibitem[{{Vesperini} {et~al.}(2018){Vesperini}, {Hong}, {Webb}, {D'Antona}, \&
  {D'Ercole}}]{Vesperini2018}
{Vesperini}, E., {Hong}, J., {Webb}, J.~J., {D'Antona}, F., \& {D'Ercole}, A.
  2018, \mnras, 476, 2731

\bibitem[{{Vesperini} {et~al.}(2013){Vesperini}, {McMillan}, {D'Antona}, \&
  {D'Ercole}}]{Vesperini2013}
{Vesperini}, E., {McMillan}, S. L.~W., {D'Antona}, F., \& {D'Ercole}, A. 2013,
  \mnras, 429, 1913

\bibitem[{{Winter} \& {Clarke}(2023)}]{winter23}
{Winter}, A.~J. \& {Clarke}, C.~J. 2023, \mnras, 521, 1646

\bibitem[{{Yaghoobi} {et~al.}(2022{\natexlab{a}}){Yaghoobi}, {Calura},
  {Rosdahl}, \& {Haghi}}]{Yaghoobi22a}
{Yaghoobi}, A., {Calura}, F., {Rosdahl}, J., \& {Haghi}, H. 2022{\natexlab{a}},
  \mnras, 510, 4330

\bibitem[{{Yaghoobi} {et~al.}(2022{\natexlab{b}}){Yaghoobi}, {Rosdahl},
  {Calura}, {Khalaj}, \& {Haghi}}]{Yaghoobi22b}
{Yaghoobi}, A., {Rosdahl}, J., {Calura}, F., {Khalaj}, P., \& {Haghi}, H.
  2022{\natexlab{b}}, \mnras, 517, 4175

\end{thebibliography}
%

\end{document}